\documentclass[useAMS,usenatbib,usegraphicx]{mn2e}
\usepackage{amsmath,amsfonts,amssymb}

\newcommand{\1}{$^{-1}$}
\newcommand{\2}{$^{-2}$}
\newcommand{\3}{$^{-3}$}
\newcommand{\hm}{$h^{-1}$}

\newcommand{\msun}{M$_{\sun}$}
\newcommand{\phot}{photon s$^{-1}$ cm$^{-2}$ sr$^{-1}$}
\newcommand{\K}{{\rm K}}
\newcommand{\kms}{{\rm km}\,{\rm s}^{-1}}
\newcommand{\ergcms}{{\rm erg}\,{\rm cm}^{-2}\,{\rm s}^{-1}}
\newcommand{\cm}{{\rm cm}}

\newcommand{\aap}{A\&A}
\newcommand{\aal}{A\&AL}
\newcommand{\apj}{ApJ}
\newcommand{\apjl}{ApJL}
\newcommand{\apjs}{ApJS}
\newcommand{\mnras}{MNRAS}
\newcommand{\pasj}{PASJ}
\newcommand{\pasp}{PASP}
\newcommand{\nat}{Nature}
\newcommand{\araa}{ARA\&A}
\newcommand{\ssr}{SSR}

\newcommand{\newar}{NewAR}
\newcommand{\exa}{ExA}
\newcommand{\spie}{SPIE}
\newcommand{\pspie}{Proc. of SPIE}

\newcommand{\ion}[2]{#1\,{\sc{#2}}}
\newcommand{\cv}{\ion{C}{V}}
\newcommand{\cvi}{\ion{C}{VI}}
\newcommand{\nvi}{\ion{N}{VI}}
\newcommand{\nvii}{\ion{N}{VII}}
\newcommand{\ovii}{\ion{O}{VII}}
\newcommand{\oviii}{\ion{O}{VIII}}
\newcommand{\neix}{\ion{Ne}{IX}}
\newcommand{\nex}{\ion{Ne}{X}}
\newcommand{\mgxii}{\ion{Mg}{XII}}
\newcommand{\sixiii}{\ion{Si}{XIII}}

\newcommand{\sxv}{\ion{S}{XV}}
\newcommand{\fexvi}{\ion{Fe}{XVI}}
\newcommand{\fexvii}{\ion{Fe}{XVII}}
\newcommand{\fexviii}{\ion{Fe}{XVIII}}
\newcommand{\fexx}{\ion{Fe}{XX}}

\newcommand{\owls}{OWLS}
\newcommand{\cloudy}{{\small CLOUDY}}
\newcommand{\chandra}{{\it Chandra}}
\newcommand{\xray}{X-ray}
\newcommand{\xmm}{{\it XMM-Newton}}
\newcommand{\ixo}{{\it IXO}}
\newcommand{\edge}{{\it EDGE}}
\newcommand{\xenia}{{\it Xenia}}
\newcommand{\genx}{{\it Generation-X}}
\newcommand{\dios}{{\it DIOS}}
\newcommand{\mbe}{{\it MBE}}
\newcommand{\estremo}{{\it ESTREMO}}
\newcommand{\fuse}{{\it FUSE}}
\newcommand{\rosat}{{\it ROSAT}}

\newcommand{\default}{\emph{REF}}                
\newcommand{\nosn}{\emph{NOSN}}                  
\newcommand{\zcool}{\emph{NOZCOOL}}              
\newcommand{\wmom}{\emph{WVCIRC}}                
\newcommand{\agn}{\emph{AGN}}
\newcommand{\mill}{\emph{MILL}}
\newcommand{\wdens}{\emph{WDENS}}
\newcommand{\wml}{\emph{WML4}}
\newcommand{\dblimf}{\emph{DBLIMF}}

\title[Soft X-rays from the WHIM]{Metal-line emission from the warm-hot intergalactic medium: I. Soft X-rays}
\author[Bertone et al.]{Serena Bertone$^{1}$\thanks{E-mail: serena@scipp.ucsc.edu},
Joop Schaye$^{2}$,
Claudio Dalla Vecchia$^{2,3}$,
C. M. Booth$^{2}$, \newauthor
Tom Theuns$^{4,5}$ and 
Robert P.~C. Wiersma$^{2,6}$
\\
$^{1}$Santa Cruz Institute for Particle Physics, University of California, 1156 High Street, Santa Cruz CA 95064, USA \\
$^{2}$Leiden Observatory, Leiden University, P.O. Box 9513, 2300 RA Leiden, The Netherlands \\
$^{3}$Max Planck Institute for Extraterrestrial Physics, Giessenbachstrasse 1, 85748 Garching bei M\"unchen, Germany \\
$^{4}$Institute for Computational Cosmology, Department of Physics, University of Durham, South Road, Durham, DH1 3LE \\
$^{5}$Universiteit Antwerpen, Campus Groenenborger, Groenenborgerlaan
171, B-2020 Antwerpen, Belgium \\
$^{6}$Max Planck Institut f\"ur Astrophysik, Karl Schwarzschild Str. 1, Postfach 1317, D-85741, Garching bei M\"unchen, Germany
}

\begin{document}
\date{Submitted to MNRAS}
\pagerange{\pageref{firstpage}--\pageref{lastpage}} \pubyear{2009}
\maketitle
\label{firstpage}

\begin{abstract}
Emission lines from metals offer one of the most promising ways to detect the elusive warm-hot intergalactic medium (WHIM; $10^5\,\K <T<10^7\,\K$), which is thought to contain a substantial fraction of the baryons in the low-redshift Universe. We present predictions for the soft X-ray line emission from the WHIM using a subset of cosmological simulations from the OverWhelmingly Large Simulations (\owls) project. We use the \owls\ models to test the dependence of
the predicted emission on a range of physical prescriptions, such as cosmology, 
gas cooling and feedback from star formation and accreting black holes. Provided that metal-line cooling is taken into account, the models give surprisingly similar results, indicating that the predictions are robust. Soft X-ray lines trace the hotter part of the WHIM ($T\ga 10^6\,\K$). We find that the \oviii\ 18.97 \AA\ is the strongest emission line, with a predicted maximum surface brightness of $\sim 10^2\,$\phot, but a number of other lines are only slightly weaker. All lines show a strong correlation between the intensity of the observed flux and the density and metallicity of the gas responsible for the emission. On the other hand, the potentially detectable emission consistently corresponds to the temperature at which the emissivity of the electronic transition peaks. The emission traces neither the baryonic nor the metal mass. In particular, the emission that is potentially detectable with proposed missions, traces overdense ($\rho \ga 10^2\,\rho_{\rm mean}$) and metal-rich ($Z\ga 10^{-1}\,Z_{\odot}$) gas in and around galaxies and groups. While soft X-ray line emission is therefore not a promising route to close the baryon budget, it does offer the exciting possibility to image the gas accreting onto and flowing out of galaxies. 
\end{abstract}


\section{Introduction}
\label{intro}

Observations of the low-redshift Universe are unable to
account for a large fraction of the baryons predicted by cosmic
microwave background (CMB) and nucleosynthesis measurements
(e.g.\ \citealt{persic1992}; \citealt{fukugita1998};
\citealt{fukugita2004}). Cosmological simulations predict that most of
these elusive baryons reside in a diffuse, warm-hot intergalactic medium
(WHIM, hereafter) with temperatures in the range $10^5\,\K
<T<10^7\,\K$ (\citealt{Cen1999}; \citealt{dave2001};
\citealt{Cen2006}; see \citealt{bertone2008} for a recent review).
Theoretical models agree with numerical simulations that the
relatively high temperature of the WHIM is produced by gravitational
heating (\citealt{sunyaev1972}; \citealt{Nath2001};
\citealt{furlanettoloeb2004}; \citealt{rasera2006}). When large-scale
structures form by gravitational collapse, cosmological shock waves
are driven into the intergalactic medium (IGM) and heat
the gas to temperatures much greater than $10^4\,\K$.

The detection of the WHIM in the low-redshift Universe
requires observations in the ultraviolet (UV) and in the soft
\xray\ bands (see e.g.\ \citealt{bregman2007} and
\citealt{prochaska2008} for reviews). 
Unfortunately, UV and \xray\ absorption lines are more
technically challenging to detect than optical lines, and require
satellites outside the Earth's atmosphere. In addition, the low density of the WHIM implies that its
surface brightness is extremely low, making it nearly impossible to
detect in emission with current instruments such as \chandra\ and
\xmm\ and the Far Ultraviolet Spectroscopic Explorer
\fuse\  (\citealt{pierre2000}; \citealt{kravtsov2002}; \citealt{Yoshikawa2003}; \citealt{furlanetto2004}; \citealt{Yoshikawa2004}; \citealt{fang2005}; \citealt{Yoshida2005}; \citealt{Yoshikawa2006}; \citealt{ursino2006}).
However, \citet{werner2008} claim to have detected continuum
emission with \xmm\ from a filament extending between the two large
clusters Abell 222 and Abell 223 at $5\sigma$ significance, thanks to
the orientation of the filament along the line of sight. \citet{zappacosta2002} and \citet{zappacosta2005} also found evidence for diffuse thermal emission from WHIM gas in two \rosat\ fields, corresponding to an overdensity of galaxies at $z\sim 0.4$ \citep{mannucci2007} and to the Scultor Wall.

Absorption lines are less biased towards high gas densities and may
therefore be more sensitive to the low-density WHIM.
There are a few controversial claims of detection of \ovii\ absorption
lines in the spectrum of Mrk 421 (\citealt{nicastro2005}; but see
\citealt{kaastra2006}; \citealt{rasmussen2007}) and from the Sculptor
Wall \citep{buote2009}. Future satellites such as the International
\xray\ Observatory\footnote{http://ixo.gsfc.nasa.gov/}
(\ixo\ hereafter, see also \citealt{markevitch2009} and \citealt{bregman2009}) and \xenia\footnote{http://sms.msfc.nasa.gov/xenia/}
\citep{hartmann2009} will be ideal to detect a large number of WHIM 
absorption lines in the spectra of QSO and gamma-ray bursts,
respectively (e.g.\ \citealt{hellsten1998}; \citealt{perna1998}; 
\citealt{kravtsov2002}; \citealt{chen2003}; \citealt{CenFang2006};
\citealt{Kawahara2006}; \citealt{paerels2008}; \citealt{branchini2009}).

Although \xray\ absorption lines are a powerful tool to investigate
the physical properties of the WHIM, the information they yield has
the limitation of being 1-dimensional, unless close quasar pairs can
be found in large numbers.  While absorption lines are not ideal to
reconstruct the 3-dimensional distribution of the gas, emission lines
do just that: they effectively map the 2-dimensional distribution of
the WHIM on the sky, and by means of their spectral redshift they
trace the distribution of the gas along the 3rd dimension. Detecting
emission from the WHIM is likely to be much more challenging than
detecting absorption, but the reward is comparatively large. It is
therefore paramount to predict accurately what level of emission can
be expected, in order to build instruments and satellites adequate to
the task.

The spatial distribution of metals in the IGM, as traced by the WHIM
emission, also provides important clues for understanding the metal
enrichment history of the Universe. The distance travelled by metals
between the sites of star formation and the lower density IGM can
constrain the energy budget of processes such as galactic
winds and AGN feedback. 


In this work, we use a subset of runs taken from the Over-Whelmingly
Large Simulations (\owls, hereafter) project \citep{schaye2010} to
investigate which 
emission lines in the soft \xray\ band best trace the low-redshift WHIM
emission. We will do the same for UV emission in a companion paper
(Bertone et al., in preparation). We estimate the level of the
emission in a simulated cubic 
region of 100 \hm\ comoving Mpc on a side and discuss the observability of the
WHIM by future telescopes such as \ixo, the Explorer of Diffuse
emission and Gamma-ray burst Explosions (\edge, \citealt{piro2009})
and \xenia.

The \owls\ runs \citep{schaye2010} are well-suited for this study and
represent a significant step forward with respect to previous
studies. \owls\ couples a large simulated dynamical range with a large
set of models in which the implementation 
of many uncertain physical processes have been varied. This special
feature allows us to investigate the dependence of the results on
different physical processes.  In this paper we will focus on the strongest
soft \xray\ emission lines we have identified, which are listed in
Table \ref{eltable}, the most important being \oviii\ $\lambda$ 18.97
\AA\ and \ovii\ $\lambda$ 21.60 \AA. When more than one line
corresponds to a given transition, we consider only the strongest
line.

\begin{table}
\centering
\caption{List of the emission lines considered in this work. The first
  column shows the name of the emission line, the second and the third
  columns the wavelength and the energy of the transition,
  respectively. The last column identifies the type of ion responsible
  for the transition (e.g.\ ``H'' means H-like) and the letter in
  parentheses 
  indicates the transition for the case of triplets from
  He-like atoms: resonance (r), forbidden (f) and intercombination
  (i).}
\begin{tabular}{l r@{}l r@{}l l}
\hline \hline Ion & $\lambda$ \; & (\AA) & E \; & (eV) & Description
\\ \hline \cv & 40. & 27 & 307.  & 88 & He (r) \\ \cvi & 33. & 74 &
367.  & 47 & H \\ \nvi & 29. & 53 & 419.  & 86 & He (r) \\ \nvii &
24. & 78 & 500.  & 24 & H \\ \ovii & 21. & 60 & 573.  & 95 & He (r)
\\ \ovii & 21. & 81 & 568.  & 74 & He (i) \\ \ovii & 22. & 10 & 560.
& 98 & He (f) \\ \oviii & 18. & 97 & 653.  & 55 & H \\ \neix & 13. &
45 & 921.  & 95 & He (r) \\ \nex & 12. & 14 & 1022. & 0 & H \\ \mgxii
& 8.  & 422 & 1471. & 8 & H \\ \sixiii & 6.  & 648 & 1864. & 9 & He
(r) \\ \sxv & 5.  & 039 & 2460. & 5 & He (r) \\ \fexvii & 17. & 05 &
726.  & 97 & Ne \\ \hline \hline
\end{tabular}
\label{eltable}
\end{table}

This paper is organised as follows. We briefly describe the \owls\
reference model and the method to calculate the line emission of the WHIM
gas in Section~\ref{owls}. Section~\ref{emission} contains predictions
for various emission lines for our reference model. Sections~\ref{angle} and
\ref{redshift} are devoted to understanding the effects of varying the
angular resolution of the detector and the redshift dependence of the
flux intensity, respectively. Section~\ref{em_from} investigates what
kind of gas produces most of the emission and Section
\ref{em_model} studies the dependence of the results on the physical model.
We discuss the prospects for observing soft \xray\ emission lines with
future space-borne telescopes in Section~\ref{detect} and we summarize
our conclusions in Section~\ref{summary}. Finally, we demonstrate in
the Appendix that our conclusions are robust with respect to changes
in both the numerical resolution and the size of the simulation
volume. We also show in the Appendix how the results vary with the
band width.

\begin{table}
\centering
\caption{Adopted solar abundances, from \citet{allende2001},
  \citet{allende2002} and \citet{holweger2001}.}
\begin{tabular}{l c | l c}
\hline \hline Element & $n_i / n_{\rm H}$ & Element & $n_i / n_{\rm
  H}$\\ \hline H & 1 & Mg & 3.47$\times 10^{-5}$ \\ He & 0.1 & Si &
3.47$\times 10^{-5}$ \\ C & 2.46$\times 10^{-4}$ & S & 1.86$\times
10^{-5}$ \\ N & 8.51$\times 10^{-5}$ & Ca & 2.29$\times 10^{-6}$ \\ O
& 4.90$\times 10^{-4}$ & Fe & 2.82$\times 10^{-4}$ \\ Ne & 1.00$\times
10^{-4}$ & & \\ \hline \hline
\end{tabular}
\label{table_abund}
\end{table}

Unless stated otherwise, we will assume the Wilkinson Microwave
Anisotropy Probe year 3 (WMAP3) $\Lambda$CDM cosmology
\citep{spergel2007} with parameters $\Omega_{\rm m}=0.238$,
$\Omega_{\rm b}=0.0418$, $\Omega_\Lambda=0.762$, $n=0.951$, and
$\sigma_8=0.74$. The Hubble constant is parameterised as $H_{\rm 0} =
100~h^{-1}\,\kms$ Mpc$^{-1}$, with $h=0.73$. The adopted solar
abundance is $Z_{\sun} =0.0127$, corresponding to the value obtained
using the default abundance set of \cloudy\ (version c07.02.02, last
described in \citealt{ferland1998}). This abundance set (see Table
\ref{table_abund}) combines the abundances of \citet{allende2001},
\citet{allende2002} and \citet{holweger2001} and assumes $n_{\rm He} /
n_{\rm H} = 0.1$. We note here that the abundances adopted by
\cloudy\ may differ by significant factors from those estimated by
\citet{lodders2003}. In particular, the abundance
of iron given by \citet{lodders2003} is almost a factor of ten smaller
than in \cloudy, that of oxygen about 20 per cent higher. This should
be kept in mind when comparing 
predictions from different studies. We stress, however, that the
assumed solar abundances play no role when we use the abundances
predicted by our simulations.

\section{The simulations}
\label{owls}

The \owls\ project \citep{schaye2010} consists of a suite of
more than fifty large, cosmological, hydrodynamical simulations that
have been produced using 
a significantly extended version of the parallel PMTree-SPH code {\sc
  gadget III} \citep[last described in][]{springel2005g}. In addition
to hydrodynamic forces, the simulations include new prescriptions for
radiative cooling, star 
formation, chemodynamics, and supernova feedback. Feedback from AGN is
included in only one of the models (see
Section~\ref{em_model}). Sections~\ref{emission} -- \ref{em_from} will
only make use of the \owls\ reference model, but we will investigate
a range of other models in Section~\ref{em_model}. 

The sub-set of the \owls\ runs used in this work have
boxes of size 100 comoving \hm\ Mpc, and contain $512^3$ particles of
both gas and dark matter. Comoving gravitational softenings are set to
$1/25$ of the mean comoving inter-particle separation, but are limited to
a maximum physical scale of 2 \hm\ kpc (which is reached at
$z=2.91$). The initial gas particle mass is $8.66\times 10^7$
\hm\ \msun. The initial conditions were generated using {\sc cmbfast}
(version 4.1; \citealt{seljak1996}) and evolved to $z=127$ using the
Zeldovich approximation from an initial glass-like state.

Radiative cooling is implemented following the prescriptions of
\cite{wiersma2009a}\footnote{Using their equation (3) rather than (4) and {\sc cloudy} version 05.07 rather than 07.02.}. In brief, net
radiative cooling rates are computed element-by-element in the
presence of the cosmic microwave background (CMB) and the \citet{haardt2001}
model for the UV/X-ray background radiation from quasars and
galaxies. The contributions of the eleven elements hydrogen, helium,
carbon, nitrogen, oxygen, neon, magnesium, silicon, sulphur, calcium,
and iron are interpolated as a function of density, temperature, and
redshift from tables that have been precomputed using the publicly
available photo-ionization package \cloudy\ \citep[last described
  by][]{ferland1998}, assuming the gas to be optically thin and in 
(photo-)ionization equilibrium.  The simulations model reionization by
\lq switching on\rq\, the \cite{haardt2001} metagalactic UV/\xray-background at $z=9$. This has the effect of rapidly heating all the gas to temperatures of $\sim 10^4\,\K$.

Star formation is treated following the prescription of
\citet{schaye2008}. Gas with densities exceeding the
critical density for the onset of the thermo-gravitational instability
($n_{\rm H}\sim 10^{-2}-10^{-1}\,\cm^{-3}$) is
expected to be multiphase and star-forming 
\citep{schaye2004}. An effective equation of 
state (EOS) is imposed with pressure $P \propto \rho^{\gamma_{{\rm eff}}}$ for densities $n_{{\rm H}} >n_{\rm H}^*=0.1~\cm^{-3}$,
normalised to $P/k=1.08\times 10^3\,\cm^{-3}\,\K$ at the 
threshold.  We use $\gamma_{{\rm eff}}=4/3$ for which both the Jeans
mass and the ratio of the Jeans length to the SPH kernel are
independent of the density, thus preventing spurious fragmentation due
to a lack of numerical resolution
\citep{schaye2008}. \citet{schaye2008} showed how the observed
Kennicutt-Schmidt law, $\dot{\Sigma}_\ast = A(\Sigma_{\rm g}/1~{\rm M}_\odot\,{\rm pc}^{-2})^n$, with $\dot{\Sigma}_\ast$ the star formation rate surface density and $\Sigma_{\rm g}$ the gas surface density, can be analytically converted into
a pressure law, which can be implemented directly into the
simulations. This has the advantage that the parameters are
observables and that the simulations reproduce the input
star formation law irrespective of the assumed equation of
state. We use the \citet{schaye2008} method,
setting $A = 1.515 \times 10^{-4}\,{\rm M}_\odot\,{\rm yr}^{-1}\,{\rm kpc}^{-2}$ and $n = 1.4$ \citep[][note that we renormalized the observed relation for a Chabrier IMF]{kennicutt1998}.

As described in \citet{wiersma2009b}, the simulations follow the timed
release of all 11 different elements that contribute significantly to
the cooling rates by massive stars (Type II
supernovae and stellar winds) and intermediate mass stars (Type Ia
supernovae and asymptotic giant branch stars), assuming a Chabrier IMF
\citep{chabrier2003} spanning the range 0.1 to 100~\msun.

Energy injection due to supernovae is included through kinetic
feedback following the prescription of
\citet{vecchia2008}. Core-collapse supernovae locally inject kinetic
energy and kick gas 
particles into winds.  The feedback is specified by two parameters:
the initial mass-loading $\eta=\dot{m}_{\rm w}/\dot{m}_*$, which
describes the initial amount of gas put into the wind, $\dot{m}_w$, as
a function of the local SFR $\dot{m}_*$, and the wind velocity $v_{\rm
  w}$.  We use $\eta=2$ and $v_{\rm w}=600~\kms$, which corresponds to
40\% of the total amount of supernova energy being available in the
form of kinetic energy and yields a cosmic star formation history in
good agreement with the observations \citep{schaye2010}.

Finally, we note that \citet{wiersma2009b} showed that the mass-weighted
mean metallicity of the WHIM is about $0.1 Z_\odot$ for the reference
model.

\subsection{Emissivity tables}
\label{tables}

\begin{figure*}
\centering \includegraphics[width=\textwidth]{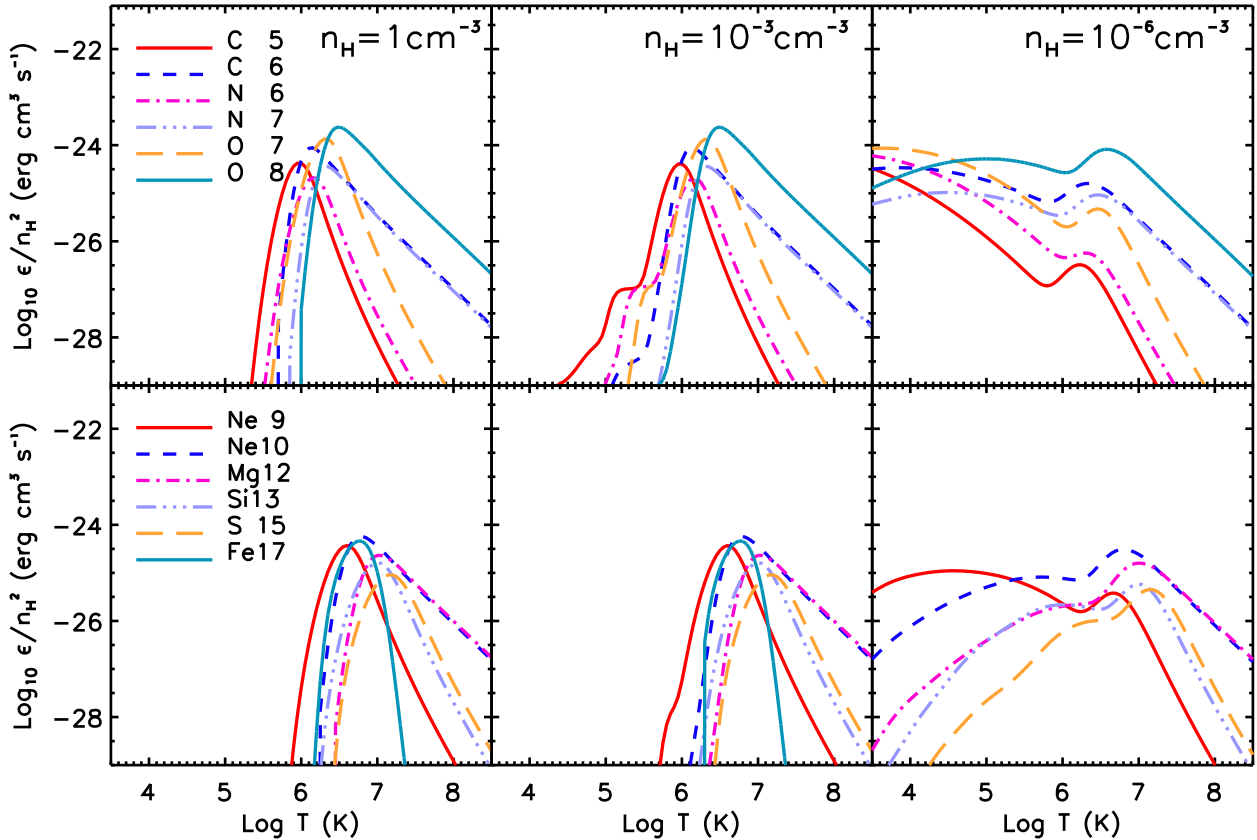}
\caption{The normalised emissivity, $\epsilon / n_{\rm
    H}^{2}$, in units of erg~cm$^3\,$s\1, of a selection of soft \xray\ emission
  lines as a function of temperature, for constant hydrogen number
  density and assuming solar abundances in the presence of the
  redshift $z=0.27$ UV/X-ray background. The left, middle, and right
  panels show results 
  for $n_{\rm H} = 1$, $10^{-3}$ and $10^{-6}\,\cm^{-3}$, respectively.
  If the gas is collisionally ionised, the normalised emissivity
  is independent of the gas density. Cool diffuse gas is mostly
  photoionised, while collisional ionisation equilibrium dominates for
  the highest temperatures and densities.} 
\label{lines}
\end{figure*}

The intensity of the emission lines is calculated by interpolating
tables of the line emissivity as a function of gas temperature,
hydrogen number density and redshift. The calculation of the emission
of individual particles in the simulation is described in Section
\ref{maps}. The tables have been created using \cloudy\ (version
c07.02.02, last described in \citealt{ferland1998} and released in
July 2008) and include about 2000 emission lines for the 11
elements tracked explicitly by our simulations.  \cloudy\ was run
using the same assumptions as were used for the calculation of the
cooling rates: optically thin gas in (photo-)ionisation equilibrium in
the presence of the CMB and the \citet{haardt2001} metagalactic
UV/X-ray background. The \cloudy\ tables were
computed assuming solar abundances (see Table \ref{table_abund}), but
are scaled to the abundances predicted by the simulation (see
Section~\ref{maps}). The 
tables have been created for the temperature range $10^2\,\K < T <
10^{8.5}\,\K$ and the hydrogen number density range $10^{-8}$ cm\3\ $<
n_{\rm H} < 10$ cm\3. The temperature is sampled in bins of $\Delta
{\rm Log}_{10}T=0.05$ and the hydrogen number density in bins of
$\Delta {\rm Log}_{10} n_{\rm H}=0.2$.

The emissivities of a selection of the strongest individual lines,
listed in Table \ref{eltable}, are shown in Fig.~\ref{lines} as a
function of temperature, for constant hydrogen number densities
$n_{\rm H} = 10^{-6}$, $10^{-3}$ and 1~cm\3 and assuming solar
abundances.  More lines are included 
in our tables, but since their intensities are much weaker than for those
listed, we do not consider them for this work. However, these weaker
lines are crucial for creating synthetic spectra of the WHIM
emission, because their cumulative contributions can significantly
increase the level of the continuum. 

Collisional ionisation is the dominant ionisation process at the
highest densities and temperatures, while the contribution of
photo-ionisation is largest at low density and at low
temperatures. The wings of the emissivity curves at $T<10^6\,\K$ and
$n_{\rm H} <10^{-3}$ cm\3\ are signatures of photo-ionisation. 
The assumption of
ionisation equilibrium, also used to calculate the cooling tables
\citep{wiersma2009a}, is well justified for photo-ionised regions and
dense gas in the centres of clusters (see \citealt{bertone2008} for a
review). However, non-equilibrium processes may become important in
the WHIM and in the outer regions of groups and clusters
(\citealt{Yoshida2005}; \citealt{Yoshikawa2006};
\citealt{CenFang2006}; \citealt{gnat2007}). \citet{gnat2009}, in
particular, have recently shown that non-equilibrium processes affect
the cooling of shock-heated gas with temperatures with $10^4\,\K
<T<10^7\,\K$, and that the amplitude of the deviations from ionisation
equilibrium increases with the gas metallicity. On the other hand,
\citet{Yoshikawa2006} found the effect of non-equilibrium ionisation
to be small for the subject of this paper: metal-line emission from
the WHIM.

The emissivities shown in Fig.~\ref{lines} are comparable to those of
\citet{Yoshikawa2003} and \citet{fang2005}, with the exception of the
\fexvii\ line, which is significantly weaker than found by
\citet{fang2005}. In general, we find that \cloudy\ predicts a
\fexvii\ emissivity a factor of about two smaller than the
Raymond-Smith code \citep{raymond1977} used by \citet{fang2005}. Part
of the reason for this discrepancy could be that two of the 
five strongest \fexvii\ emission lines are missing from the
\cloudy\ line list, namely $\lambda 15.01$ \AA\ and $\lambda 17.10$
\AA. This might be an issue for the cooling of metal-rich gas at
$T\sim 10^7\,\K$, where the \fexvii\ line can contribute up to 30 per
cent of the total cooling rate. On the other hand, this lack of iron
lines might be balanced by the high iron abundance assumed in \cloudy,
compared to that estimated by \citet{lodders2003}.

\subsection{Line emission calculations}
\label{maps}

We investigate the WHIM emission in the soft \xray\ band by creating
synthetic maps of the surface brightness of individual emission lines.
We only include emission from gas with densities $n_{\rm H}<0.1\,\cm^{-3}$, because our simulations lack the resolution
and the physics to model the emission from higher density gas, which
is expected to be multiphase \citep[see the discussion in][]{vecchia2008}. 

We calculate the luminosity of emission line $l$ for each gas particle
$i$ as a function of its gas mass $m_{{\rm gas},i}$, density $\rho_i$,
hydrogen number density $n_{{\rm H},i}$ and element abundance $X_{{\rm y},i}$ as
\begin{equation}\label{lumin}
L_{i,l}\left(z\right) = \varepsilon_{i,l,\odot} \left( z, T_{\rm i},n_{\rm H} \right) \frac{m_{{\rm gas},i}}{\rho_i} \frac{X_{{\rm y},i}}{X_{\rm y \sun}} \quad\left(\textrm{erg s}^{-1}\right),
\end{equation}
where $\varepsilon_{i,l,\odot} \left( z, T_{\rm i},n_{\rm H}\right)$ is the emissivity of the line $l$ for solar abundances, in units of erg cm\3\ s\1, bi-linearly interpolated (in logarithmic space) from the \cloudy\ tables described
in Section \ref{tables} as a function of the particle temperature
${\rm Log}_{10} T_{\rm i}$ and hydrogen number density ${\rm Log}_{10}
n_{{\rm H},i}$ at the desired redshift.  $X_{{\rm y},i}$ is the mass
fraction of element $y$ and $X_{\rm y \sun}$ its solar value. We use
the ``smoothed'' element abundances described in \citet{wiersma2009b},
which are defined as the ratio of the SPH smoothed mass density of the element
and the SPH smoothed total mass density, $Z_{{\rm y},i} = \rho_{{\rm y},i} /
\rho_{i}$. While the use of smoothed abundances, which were also used to
calculate the cooling rates during the simulations, reduces the
effects of the lack of metal mixing inherent to SPH, it does not solve
the problem (see \citealt{wiersma2009b} for a discussion).

The corresponding particle flux $F_{i,l}$ is given by
\begin{equation}
F_{i,l} = \frac{L_{i,l}}{4\pi D_{\rm L}^2} \frac{\lambda_l}{h_{\rm P}c} \left(1+z\right) \quad\left(\textrm{photon s}^{-1}\textrm{ cm}^{-2}\right),
\end{equation}
with $D_{\rm L}$ the luminosity distance, $h_{\rm P}$ the Planck
constant, $c$ the speed of light and $\lambda_l$ the rest-frame
wavelength of the emission line $l$.  Finally, we project the fluxes
on to a 2-dimensional grid whose number of pixels is determined
by the angular resolution we require and by the redshift we
want to ``observe''. We will investigate the effect of changing the
pixel size in Section~\ref{angle}. The projection is done using a flux-conserving SPH interpolation scheme.

Surface brightness ($S_{\rm B}$, in all figures) maps are computed by dividing the flux in each pixel by its solid angle $\Omega_{\rm pixel} = 2\pi \left[ 1- \cos \left(\vartheta_{\rm rad}/2 \right) \right]$, where $\vartheta_{\rm rad}$ is the pixel size in units of radians.
At $z=0.25$ our simulation box corresponds to a field of view of about 8 degrees and an angular resolution of 15" requires a grid with $1926^2$ pixels. At
$z=0.25$ this corresponds to a physical scale of about 42 \hm\ kpc,
which is about 21 times larger than the gravitational softening scale
used in the simulation.  Unless stated otherwise,
the emission maps shown were created with a
pixel size corresponding to an angular resolution of 15" on the
sky. We divide each simulated box of 100 \hm\ Mpc into 5 slices that
are each 20 \hm\ Mpc thick and create as many emission maps, which we combine to
calculate the flux probability distribution function (PDF,
hereafter) corresponding to this angular resolution and slice
thickness. We neglect the effect of peculiar velocities when creating the slices, and select particles only according to their position within the simulation box. At $z=0.25$, a slice thickness of 20 \hm\ Mpc corresponds 
to an energy resolution
\begin{equation}
\frac{\Delta E}{E} =  \frac{\Delta v}{c} = \frac{H\left(  z\right)}{c} \frac{\Delta L}{1+z} \approx 6\times 10^{-3},
\end{equation}
which is comparable to the spectral resolution proposed for \ixo\ and
\edge. In Appendix~\ref{thick} we demonstrate that the flux
probability density function is proportional to the slice
thickness, at least for $\Delta L>5~h^{-1}\,{\rm Mpc}$ and
sufficiently large fluxes (greater than $10^{-5}\,$\phot\ for \oviii\ and our
fiducial angular resolution of 15"). This is because the probability
of a pixel having a certain flux is proportional to the volume
probed, provided the slice is sufficiently thick to encompass the
emitting objects and sufficiently thin for projection effects to be
unimportant (this last condition breaks down at very low flux levels).

\section{Emission maps}
\label{emission}

\begin{figure*}
\centering \includegraphics[width=\textwidth]{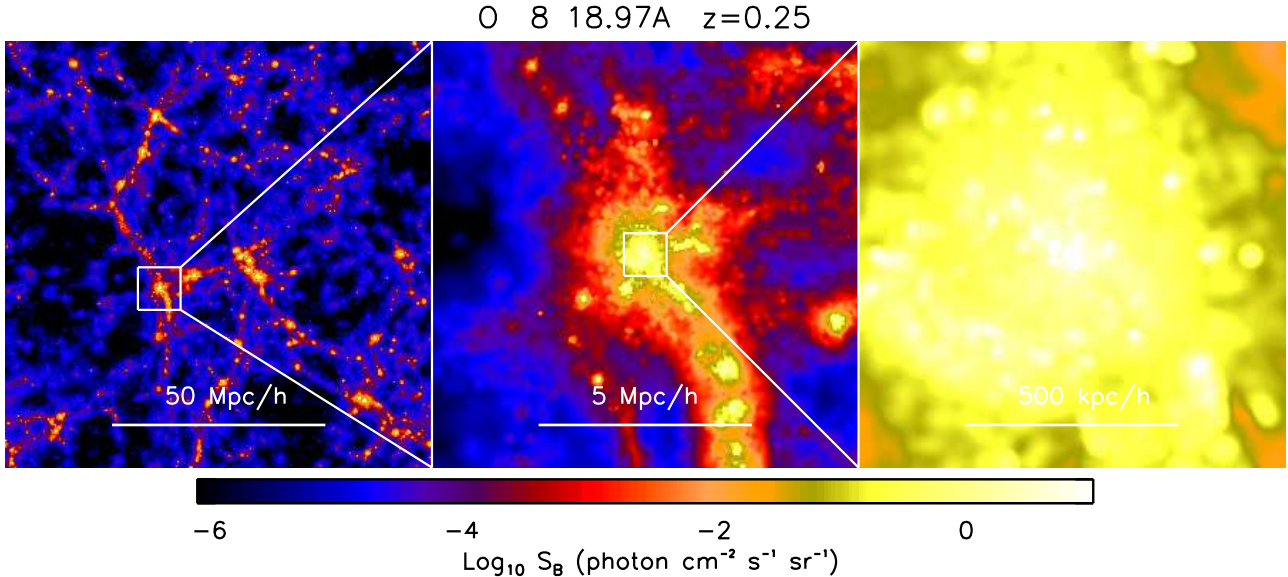}
\caption{Zoom into an \oviii\ 18.97 \AA\ emission map of a high-density region at $z=0.25$.  The
  object at the centre of the displayed region is one of the largest
  haloes in the simulation, with a total mass of a few times $10^{14}$
  \msun.  The slice thickness is 20 \hm\ comoving Mpc (which corresponds to
  $1772~\kms$). From left-to-right, the 
  pixel sizes are 100",
  10" and 1" and the angular sizes of the
  regions shown are 8 degrees, 48' and 4.8'
  (corresponding to 100, 10, and 1~\hm\ comoving Mpc), respectively. The
  blobs in the right panel are resolved substructures.}
\label{zoom_maps}
\end{figure*}

\begin{figure*}
\centering \includegraphics[width=\textwidth]{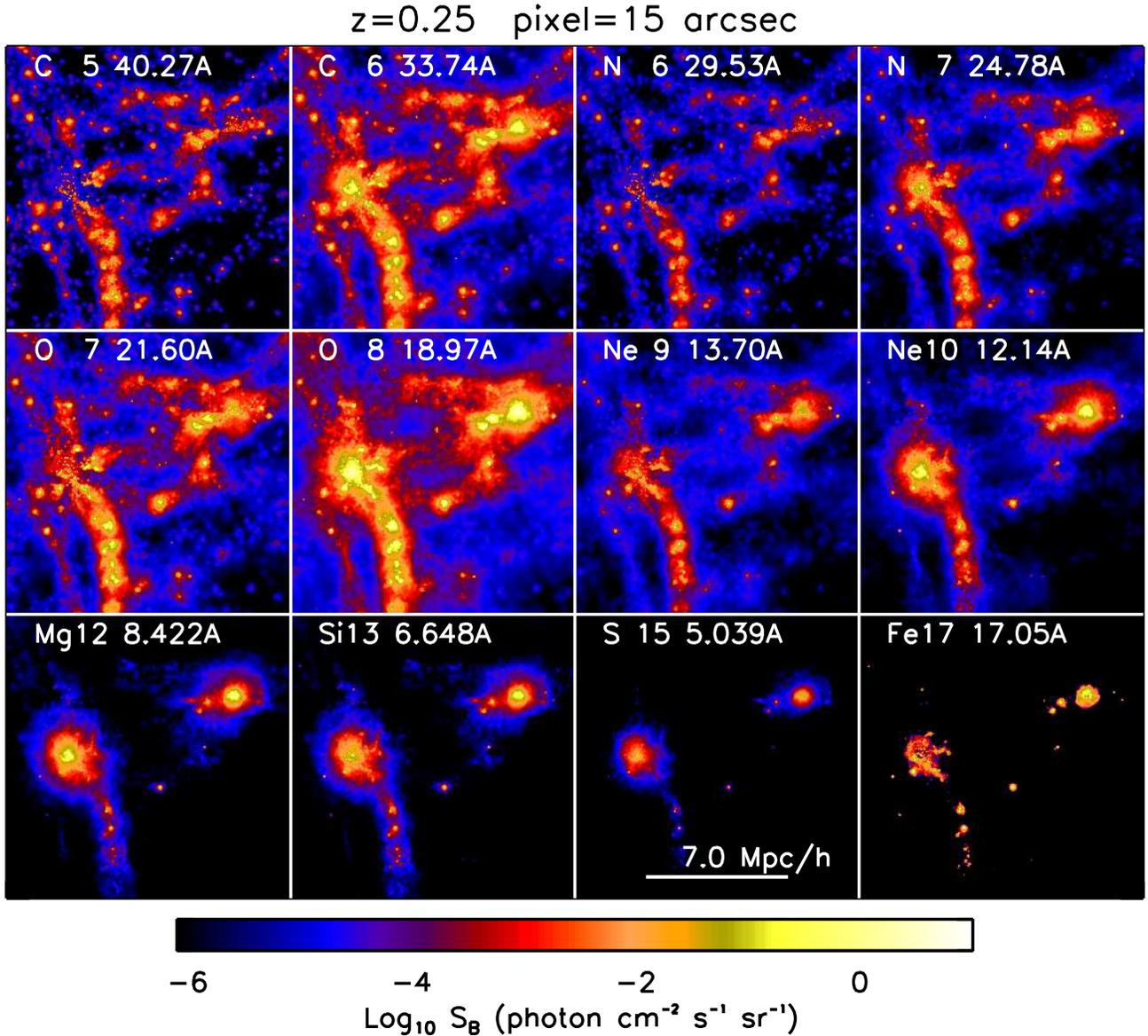}
\caption{Maps of the emission for a sample of twelve different soft
  \xray\ lines, listed in table \ref{eltable}, at $z=0.25$. The figure
  shows a zoom 
  of a region, with a comoving size of 14 \hm\ Mpc (1.12 degrees on
  the sky), that includes the largest group in the 
  simulation. The density, temperature, and metallicity in this region
  are shown in Fig.~\protect\ref{dtz}. All panels assume the same
  angular resolution (15", 
  which corresponds to a physical size of 42~\hm~kpc) and
  slice thickness (20~\hm~comoving Mpc, which corresponds to
  $1772~\kms$). The maximum of the colour scale 
  has been set to a fraction of the real maximum to enhance the
  emission of the weaker lines and of low-density regions. The
  \oviii\ line is the strongest emission line in the sample. Carbon,
  nitrogen, oxygen and neon lines best trace the distribution of the
  WHIM in filaments and in the outskirts of groups, while lines from
  elements with higher atomic numbers such as \mgxii, \sixiii,
  \sxv\ and \fexvii\ mostly trace hot and dense gas in massive haloes.}
\label{em_maps}
\end{figure*}

\begin{figure*}
\centering
\includegraphics[width=8.7cm]{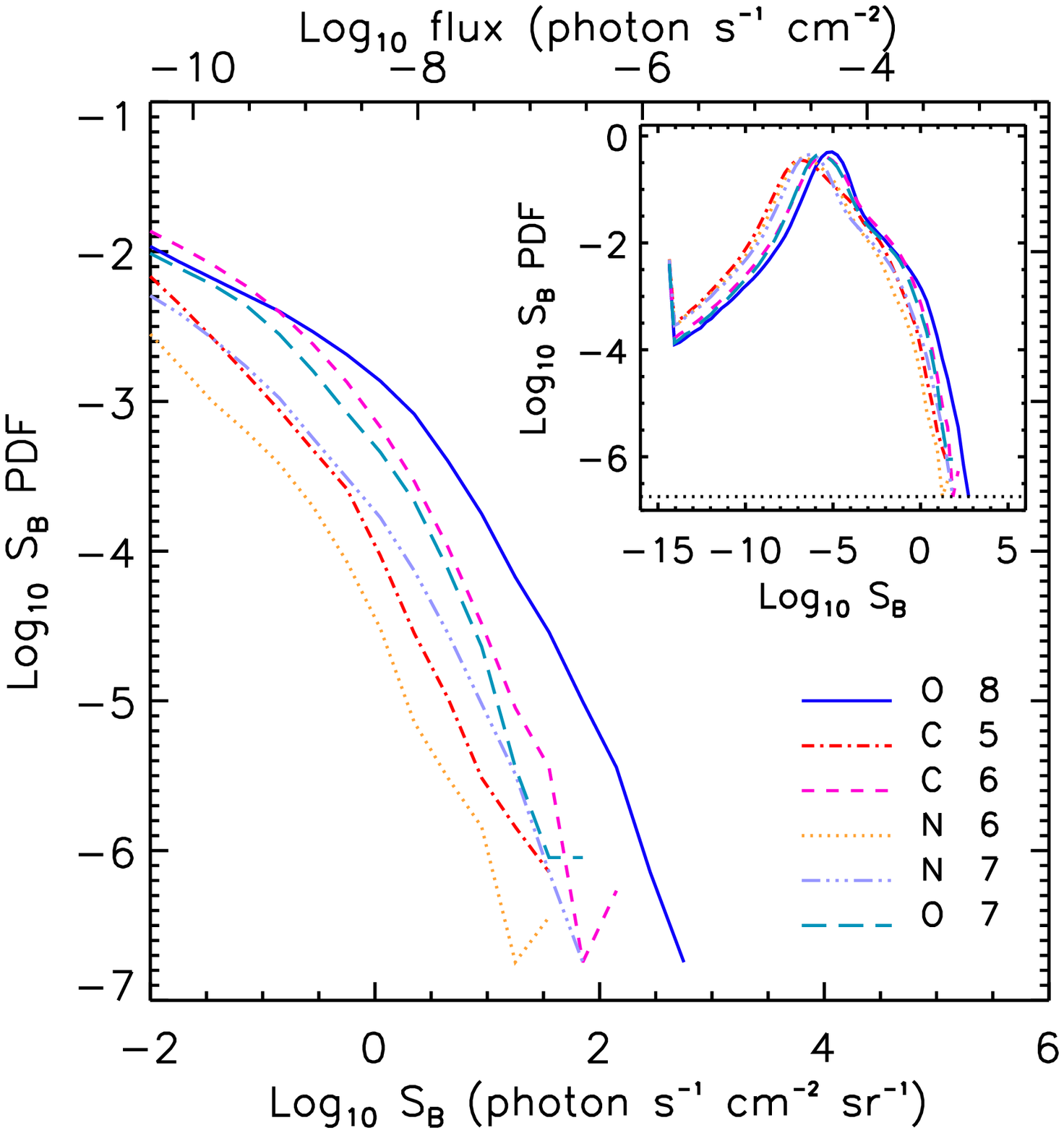}
\includegraphics[width=8.7cm]{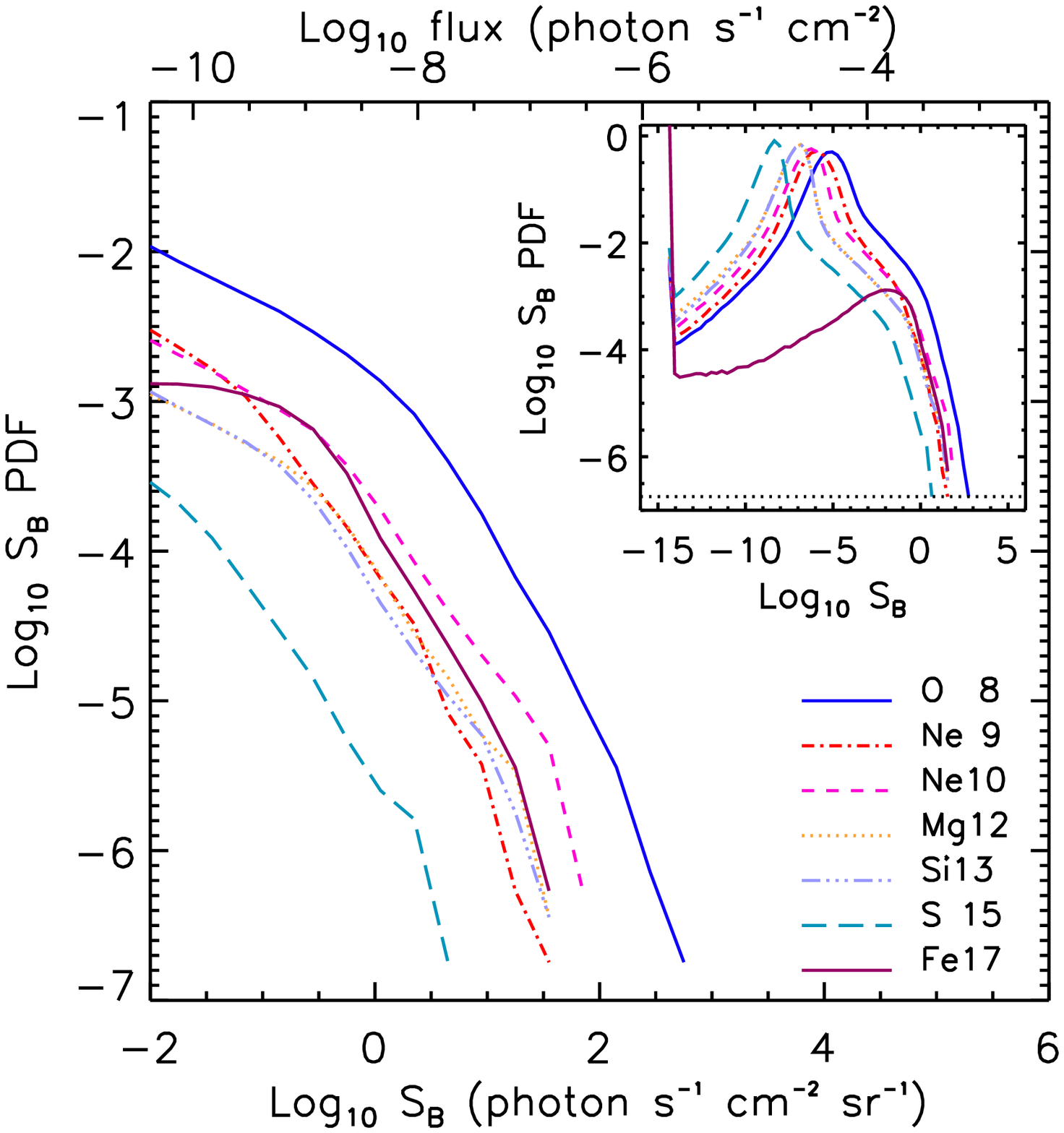}
\caption{The flux PDFs of the sample of soft \xray\ emission
  lines that is also shown in Fig.~\ref{em_maps} and that is listed in
  table~\ref{eltable}, at $z=0.25$. The left panel shows results for
  \cv, \cvi, \nvi, \nvii, \ovii\ and \oviii, and the right panel for
  \neix, \nex, \mgxii, \sixiii, \sxv\ and \fexvii.  The main plotting
  area shows only the high flux tail of the distributions,
  while the full PDFs are shown in the inset. The pixel size is 15"
  (which corresponds to a physical size of 42~\hm~kpc) and 
  PDFs are calculated using five slices through the
  simulation box that are each 20~\hm~comoving Mpc thick (which
  corresponds to $1772~\kms$). In Appendix~\ref{thick} we show that
  for sufficiently large fluxes ($>10^{-5}$~\phot\ for \oviii)
  the PDF is proportional to the slice thickness. The \oviii\ line is the
  strongest emission line in 
  the sample, followed by \cvi\ and \nex. The \fexvii\ can reach high
  fluxes, but only in a limited number of pixels compared with
  other lines (see the inset of the right panel). This is due to the different
  spatial distribution of 
  iron in the IGM and to the sharply declining values of the
  \fexvii\ emissivity curve outside a narrow range around its peak
  temperature. The intensity of the unresolved \xray\ background ($\sim 10^{-2}\,$\phot, discussed in Section \ref{detect}) corresponds to the minimum value shown on the $x$-axis.}
\label{em_pdf}
\end{figure*}

\begin{figure*}
\centering
\includegraphics[width=8.7cm]{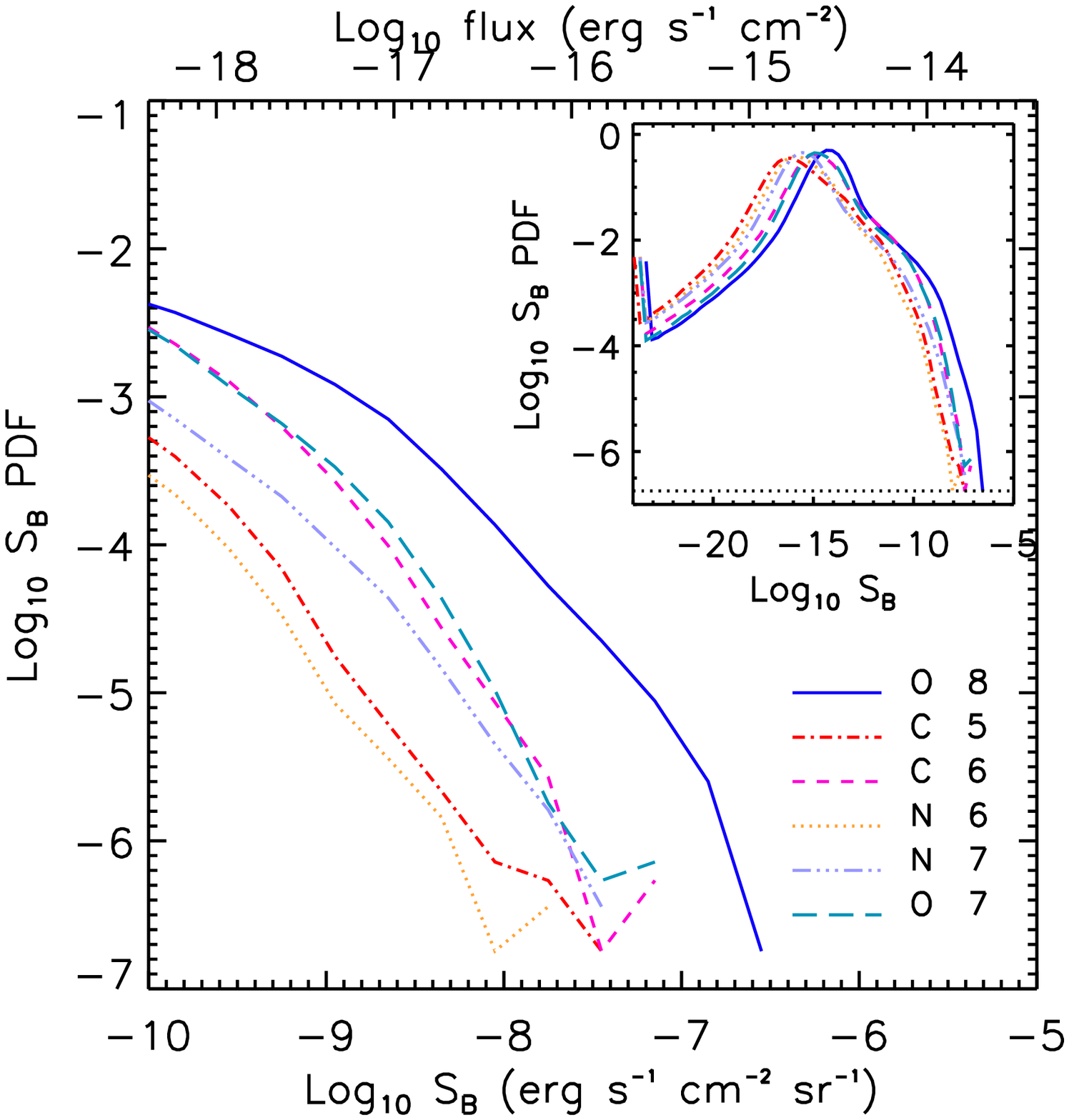}
\includegraphics[width=8.7cm]{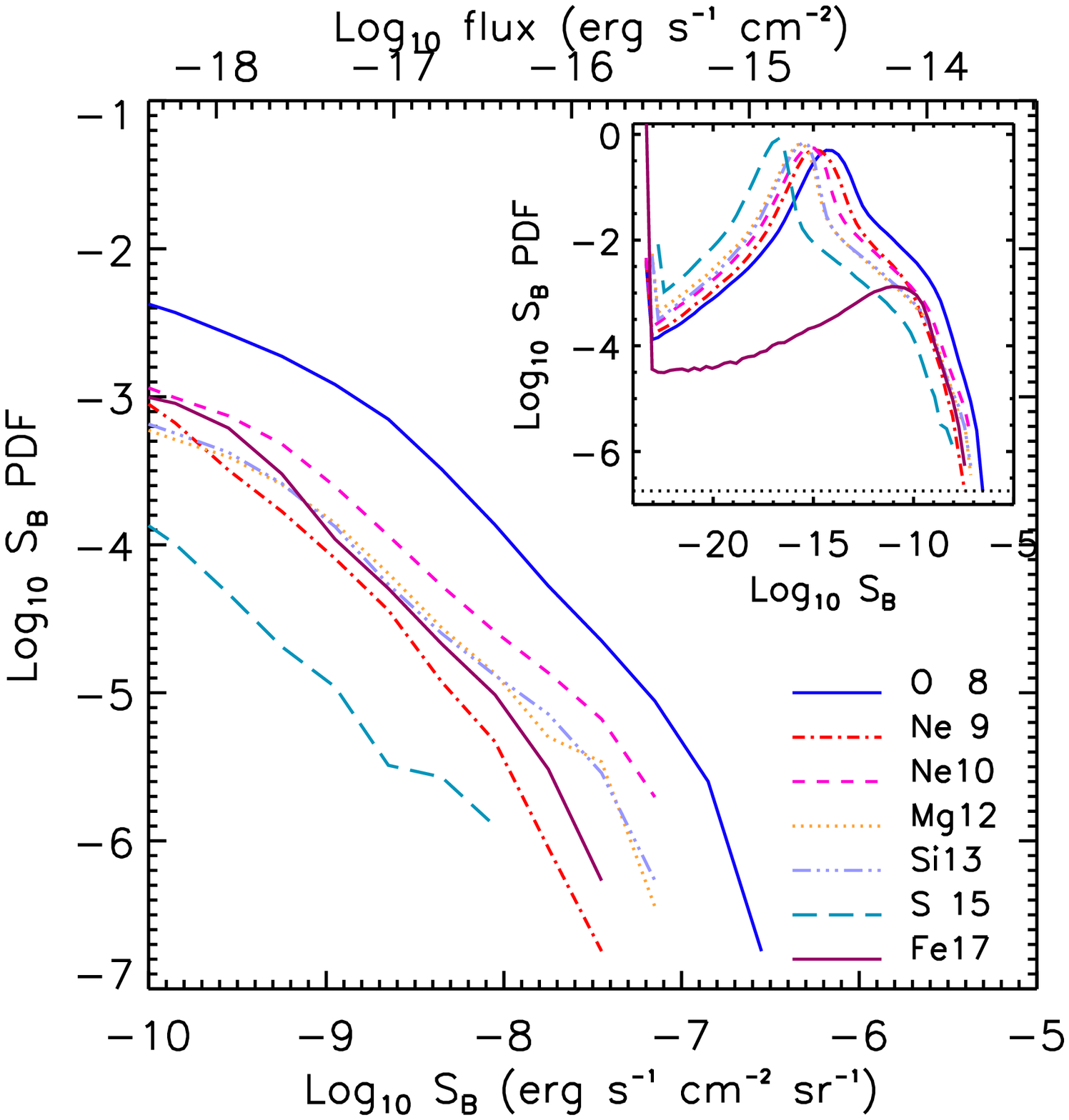}
\caption{As Fig. \ref{em_pdf}, but with the emission line flux expressed in erg s\1\ cm\2\ sr\1, instead of \phot. The relative strength of the lines varies slightly when converted to these units and is more easily comparable to the flux limits proposed for future missions such as \ixo\ and \edge.}
\label{em_pdf2}
\end{figure*}

\begin{figure*}
\centering
\includegraphics[width=\textwidth]{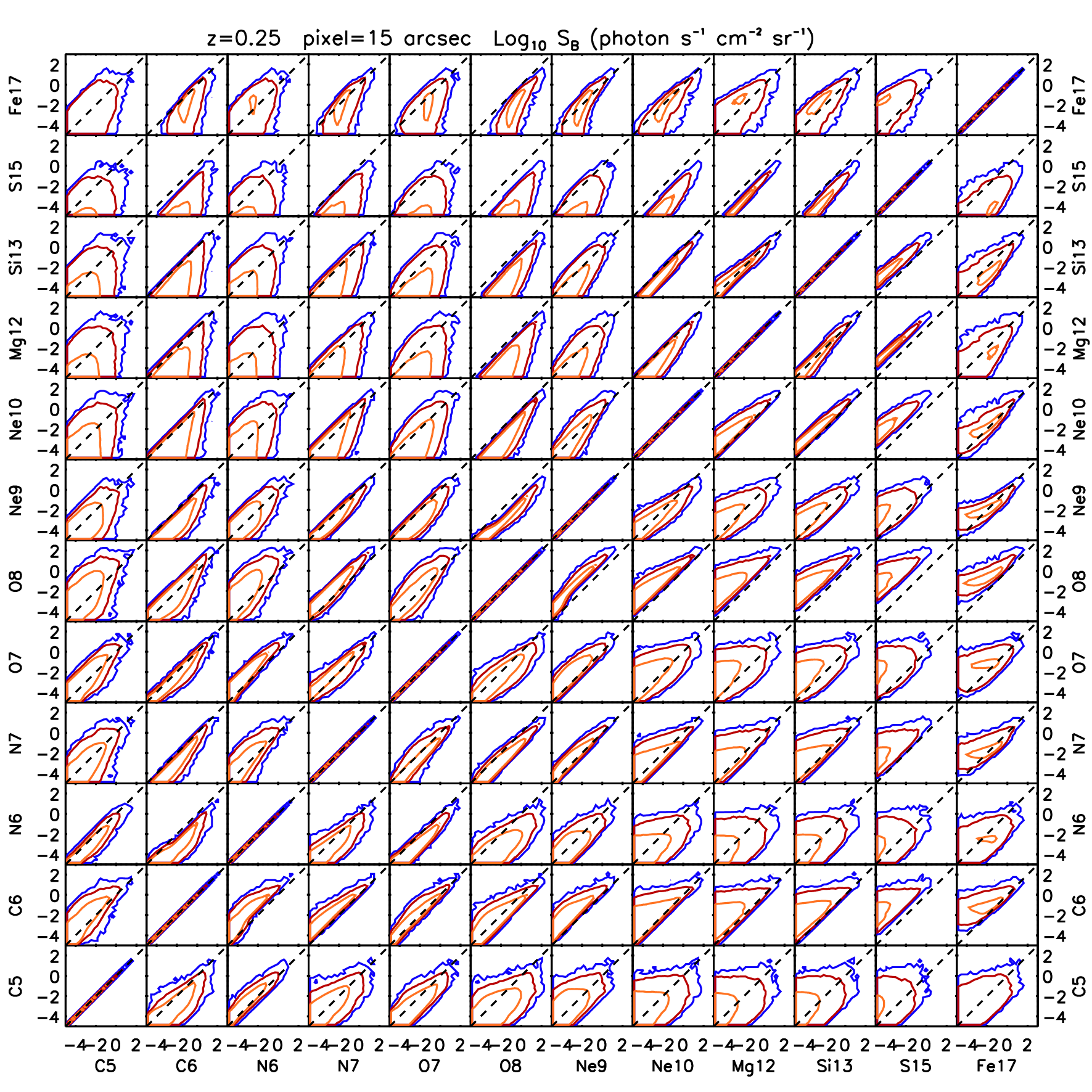}
\caption{Comparison of the intensities of the lines listed in
  Table~\ref{eltable}. The contours show the number density of pixels
  and are logarithmically spaced by 1.7~dex. The distributions have
  been calculated using five 20 \hm\ comoving Mpc slices through the
  simulation box at $z=0.25$, assuming an angular resolution of
  15". Both axes show the line surface brightness ${\rm Log}_{10}
  S_{\rm B}$ in units of \phot. The dashed diagonal line in each panel
  indicates equal fluxes. The \oviii\ and \cvi\ lines produce 
the highest fluxes. Lines whose emissivities peak at similar
temperatures (i.e.\ panels near the diagonal running from the
bottom-left to the top-right of the plot) are strongly correlated.} 
\label{fluxzz}
\end{figure*}

In this Section we present emission maps for the strongest soft
\xray\ emission lines we have identified as WHIM tracers. 
All PDFs are calculated using the full simulation box with size 100
\hm\ Mpc. However, for clarity we show
emission maps for only a fraction of the full
box. In particular, we focus on a smaller area of 14 \hm\ Mpc on a
side that contains one of the largest groups in the simulation, with a
total mass of about $7\times 10^{14}$ \msun, a smaller group and a few
dense filaments. This choice is motivated by the fact that the
strongest emission comes from high-density regions such as groups and
filaments.

To illustrate the morphology of the emission, 
Fig.~\ref{zoom_maps} shows the \oviii\ emission from a slice of the full
simulation box in the left panel and two consecutive zooms into the core of
a massive group in the companion panels. The displayed regions
decrease in size by factors of ten, ranging from 100 \hm\ Mpc to 1
\hm\ Mpc. The strongest \oviii\ emission is concentrated in the
high-density knots at the intersections of filaments, where groups
reside, while the emission from cooler and more diffuse gas in filaments and
voids is several orders of magnitude weaker. For comparison, the intensity of the unresolved \xray\ background is $\sim 10^{-2}$ \phot\ (\citealt{hickox2009}, see also Section \ref{detect}) and roughly corresponds to orange in the colour scale. As we will see in the following, the surface brightness of the WHIM is higher than the background only in relatively overdense regions.

Fig.~\ref{em_maps} shows emission maps for a selection of twelve of
the strongest emission lines in the soft \xray\ band, listed in Table
\ref{eltable}. The surface brightness PDFs for the same
lines are shown in Fig.~\ref{em_pdf} in units of \phot, and in Fig.~\ref{em_pdf2} in units of erg s\1\ cm\2\ sr\1.
To convert a flux $F_{\rm ph}$ in units of \phot\ to a flux $F_{\rm erg}$ in units of $\ergcms$, we use the conversion
\begin{equation}
F_{\rm erg} = F_{\rm ph} \frac{h_{\rm p} c}{\lambda_0\left( 1+z\right)},
\end{equation}
with $\lambda_0$ the rest-frame wavelength of the transition (e.g.\ 18.97 \AA\ for \oviii\ emission).

Here we consider only the strongest lines from Hydrogen-like atoms and the resonance lines in triplets from Helium-like atoms, and we defer the discussion of the behaviour of lines within the triplet to Section \ref{o7}.  These
lines have been chosen on the basis of their strength and
detectability in the soft \xray\ band, but are by no means the only
important lines.  Although we only show results for one iron line
from the \fexvii\ ion, we note that the soft \xray\ band is a very rich band for
iron lines from a multiplicity of other higher-order ions, including,
e.g., \fexviii\ and \fexx. Most lines from these ions trace slightly
hotter gas than
the \fexvii\ lines, but nevertheless represent an extremely important
cooling channel for hot gas in dense environments. Lines from
\fexvi\ are instead emitted at longer wavelengths and fall outside the
soft \xray\ band.

Figs. \ref{em_maps}, \ref{em_pdf} and \ref{em_pdf2} clearly indicate that
\oviii\ $\lambda 18.97$ \AA\ is the strongest emission line and that
for all lines the strongest fluxes are concentrated in very small
regions or, equivalently, in a small fraction of the pixels in the
maps. All lines are weaker than \oviii\ by a factor of a few or
more. This is a consequence of the \oviii\ $\lambda 18.97$ \AA\ line
emissivity, which is highest of all lines, and of the high relative
abundance of oxygen. The emissivity of \oviii\ peaks at a temperature
$T\sim 3\times 10^6\,\K$, and remains higher than for any other line at
higher temperatures.  The line emissivities shown in Fig.~\ref{lines}
give a good indication of what lines dominate the emission. However,
the total emission in each line is not only determined by the
emissivity itself, but also by the fraction of the gas mass and
metallicity in the temperature-density plane that corresponds to each
emissivity value.

Fig.~\ref{em_maps} illustrates how some lines, namely \cv, \cvi,
\ovii\ and \neix, preferentially trace the cooler gas in filaments and
in the outskirts of groups, while others, namely \nex, \mgxii,
\sixiii\ and \sxv\ better trace the hot medium at the centres of
groups. This is a direct consequence of the fact that the peak of the
emissivity shifts to higher temperatures with
increasing atomic number and ionisation state, as can be seen in
Fig.~\ref{lines}.  The \cv, \ovii\ and \neix\ lines, whose
emissivities peak at about $10^6\,\K$, $2\times 10^6\,\K$ and $4\times
10^6\,\K$, respectively, are best-suited to investigate the diffuse gas
in filaments and in the outskirts of groups. However, their emission
declines drastically in the centres of large haloes, where the gas is
very hot and more highly ionised atoms dominate (see e.g.\ the centres
of the groups shown in Fig.~\ref{em_maps}). Emission lines such
as \cvi, \oviii, \nex, \mgxii, \sixiii\ and \sxv\ are the best probes
of the dense gas in groups and clusters, but do not trace the low
density gas in filaments and the field as well as \cv, \ovii\ and
\neix\ do, because their emissivities peak at higher temperatures.

The \fexvii\ line is an exception to this trend because, even though
its emissivity peaks at $T\sim 6\times 10^6\,\K$, its value is only high
in a narrow temperature range. In addition, the spatial abundance of
iron differs significantly from that of other elements because of the
delay in the SN Ia enrichment. While iron is abundant in the
intragroup medium and in the vicinity of galaxies, it is significantly
underabundant in the diffuse IGM.  The flux PDFs in Fig.~\ref{em_pdf}
show that most lines produce fluxes higher than about 1 \phot\ in at
least a few pixels.

Fig.~\ref{fluxzz} shows the relative intensity of lines with respect
to each other. When the contours lie above the dashed diagonal line,
the emission line
on the $y$-axis produces the highest flux, when below, the line on the
$x$-axis is strongest. In general, the \oviii\ and \cvi\ lines produce
the highest fluxes, both among the lines that trace the hottest gas and among
those that trace cooler gas. Lines from helium-like atoms produce
higher fluxes than 
\oviii\ and \cvi\ in a relatively small number of pixels. These
correspond to regions of the IGM with relatively low temperatures,
where helium-like atoms are the dominant species. The lines in
Fig.~\ref{fluxzz} are sorted by atomic mass and then by ionisation
state, which is nearly the same as sorting by the temperature for
which the emissivity peaks (see Fig.~\ref{lines}). This explains
why the contours are much narrower for panels near the diagonal
running from the bottom-left to the top-right of the 
figure than they are for panels near the top-left and bottom-right of
the plot: panels near the diagonal correlate lines for which the
emissivity peaks at similar temperatures. Fig.~\ref{fluxzz} illustrates
the distribution of line ratios that can be 
expected in surveys of WHIM emission and may
help to confirm the validity of line identifications. 

Although the actual difference in flux between emission lines can be
as high as an order of magnitude, all emission lines in our set might
be detectable in the centres of galaxy groups and in the outskirts of
clusters.

\subsection{The \ovii\ triplet}
\label{o7}

\begin{figure*}
\centering \includegraphics[width=\textwidth]{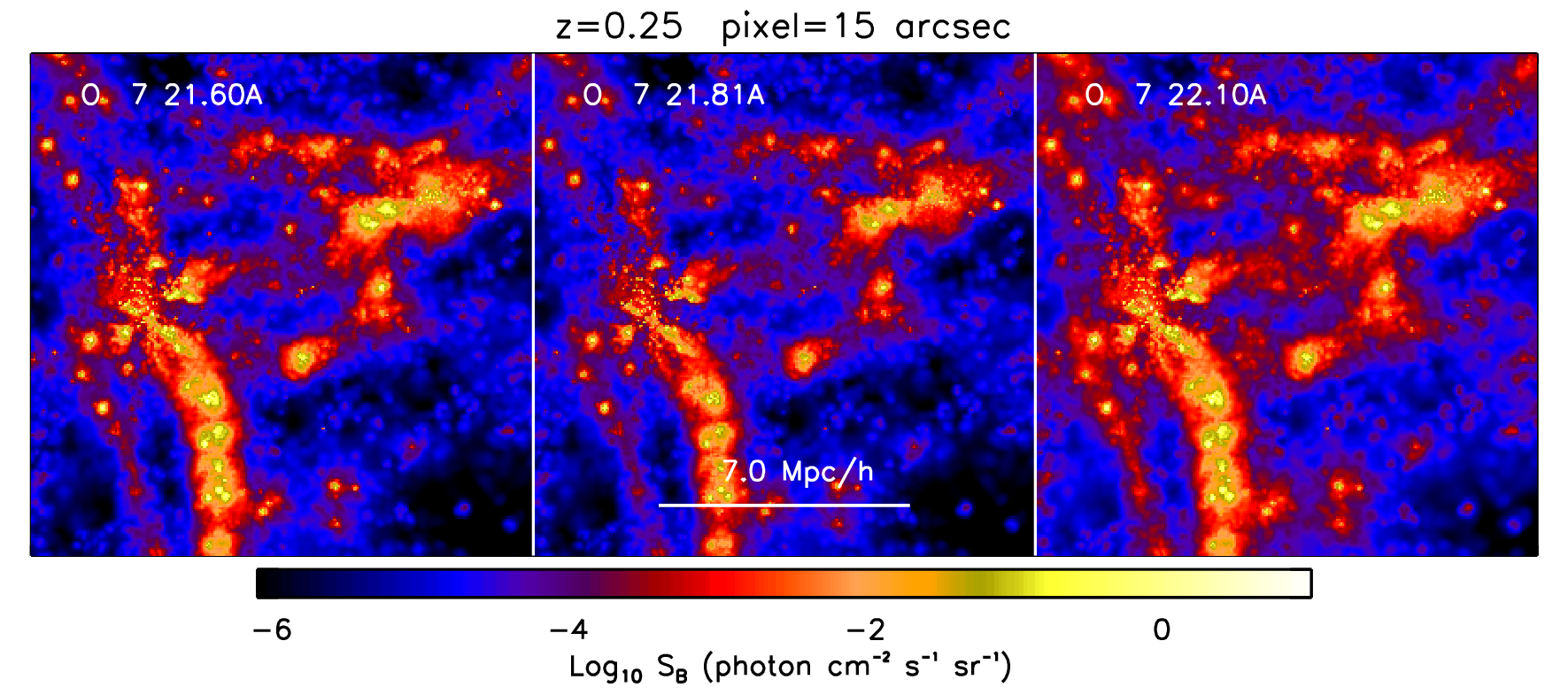}
\caption{As Fig.~\protect\ref{em_maps}, but comparing the resonance (left panel), intercombination (middle panel) and forbidden (right panel) emission
  lines of the \ovii\ triplet. The resonance and forbidden lines have similar
  strengths in dense regions with strong emission, but the forbidden
  line becomes the main emission channel in lower density regions with
  fluxes lower than about $10^{-3}$ \phot.}
\label{o7maps}
\end{figure*}

\begin{figure*}
\centering \includegraphics[width=\textwidth]{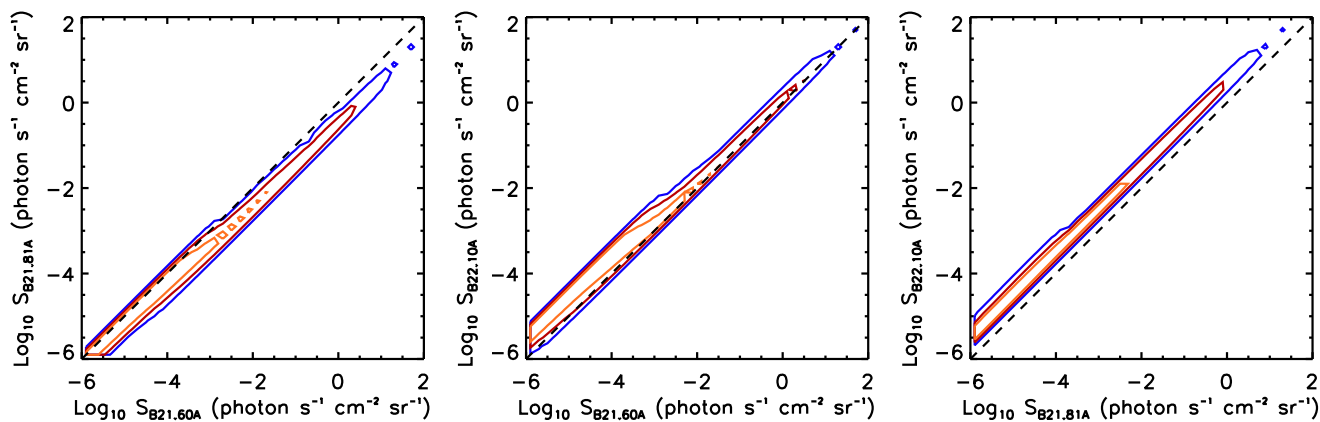}
\caption{As Fig. \ref{fluxzz}, but for the line intensities in the
  \ovii\ triplet: intercombination line vs. resonance (left panel),
  forbidden vs. resonance (middle panel) and forbidden
  vs. intercombination (right panel).}  
\label{o7ratio}
\end{figure*}

\begin{figure}
\centering \includegraphics[width=8.4cm]{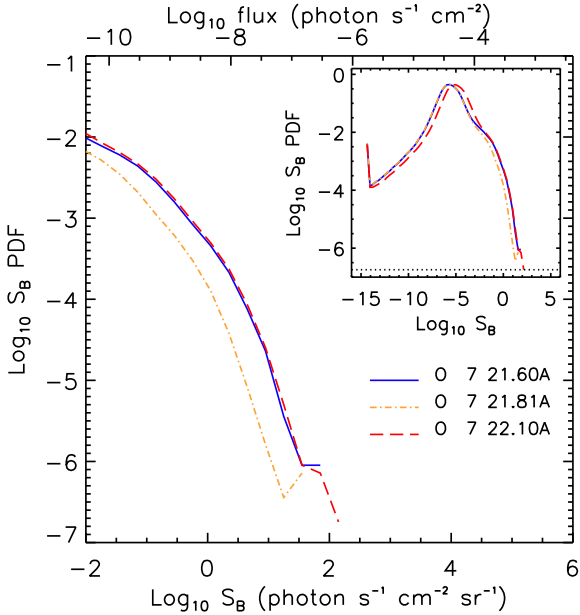}
\caption{As Fig.~\protect\ref{em_pdf}, but for the flux
  PDFs of the \ovii\ triplet emission 
  lines. The fluxes from
  the resonance and forbidden lines are about equal for ${\rm
    Log}_{10} F > -3$ \phot, but at lower fluxes the forbidden line
  becomes the dominant emission channel, while the intercombination
  line becomes as strong as the resonance line.}
\label{o7pdf}
\end{figure}

In this Section we discuss the properties of line triplets from
Helium-like atoms, focussing in particular on the \ovii\ triplet
(see Table \ref{eltable}), which should be the easiest to detect.  The
properties of the lines in triplets are important because the column
density of a gas can be estimated using the ratio of the intensity of
the resonance line of Helium-like atoms and other lines in the
triplet. This might be possible, for example, when at least two out
of the three lines are detected.

The lines in the triplet correspond to the atomic transition $1s2p \rightarrow 1s^2$ and in particular\footnote{Ralchenko Y., Kramida A.E., Reader J., and NIST ASD Team (2008). NIST Atomic Spectra Database (version 3.1.5). Available: http://physics.nist.gov/asd3. National Institute of Standards and Technology, Gaithersburg, MD.}:
\begin{enumerate}
\item resonance line: $^1P_1\rightarrow ^1S_0$
\item intercombination line: $^3P_{2,3}\rightarrow ^1S_0$
\item forbidden line: $^3S_1 \rightarrow ^1S_0$
\end{enumerate}

Of the three main emission lines of Helium-like atoms, the resonance
line is usually of comparable strength to the forbidden line at high
density, while the intercombination line is usually the weakest. The
forbidden line, however, becomes the dominant emission line in low
density environments, where gas is mostly photo-ionised and radiative
intercombination is the dominant line emission process.

Photons at the wavelength of the resonance line can be resonantly
scattered in random directions when passing through an intervening
medium, while photons at the intercombination and forbidden
frequencies have negligible cross-sections. As such, only the
intensity of the resonance line can be attenuated along the line of
sight, giving rise to absorption in the spectra of background objects.
Interestingly, as \citet{churazov2001} point out, the photons of
the cosmic \xray\ background are scattered in the same way, therefore
boosting the line emission of the WHIM gas in filaments where the
scattering occurs, compared to the thermal emission from the gas
itself. This effect might improve the chances of detecting the WHIM in
emission \citep{Yoshikawa2003}.

In Fig.~\ref{o7maps} we show the flux predicted for the resonance,
intercombination and forbidden lines in the \ovii\ triplet. The flux
PDF of the same lines is shown in Fig.~\ref{o7pdf}, while
Fig.~\ref{o7ratio} compares the flux intensity of pairs of lines on a
pixel by pixel basis.  The intensity of the intercombination line is
about 2-10 times smaller than those of the forbidden and resonant
lines. At the highest fluxes, $F \geq 10^{-3}$ \phot, corresponding to
high-density regions, as we will show in the next Sections, the
intensity of the resonance line matches that of the forbidden line
well, and the shapes of the PDFs are basically the same. At lower densities,
the fluxes predicted for the forbidden line become higher than those
of the resonance line by up to a factor of 10. This can be seen, for
example, as a blue glow in voids in the right panel of
Fig.~\ref{o7maps}, while the flux of the resonance line in the left
panel is lower than the minimum of the colour scale. As a consequence,
the ratio of the intensity of the resonance and the
intercombination line, as well as the ratio of the resonance and
the forbidden line, varies (left and middle
panels of Fig.~\ref{o7ratio}), while only the ratio of the
intercombination and forbidden lines is constant (right panel of
Fig.~\ref{o7ratio}).

In the absence of absorption by an intervening medium, both the
forbidden and the resonance lines might be used to investigate the
properties of WHIM gas. However, in low-density regions such as
filaments and in the presence of an optically thick layer of gas, the
forbidden line might be a better tracer of the WHIM. On the other
hand, the resonance line is the only useful line for
comparisons with absorption studies, such as those planned for \ixo.

\section{Angular resolution}
\label{angle}

\begin{figure}
\centering \includegraphics[width=8.4cm]{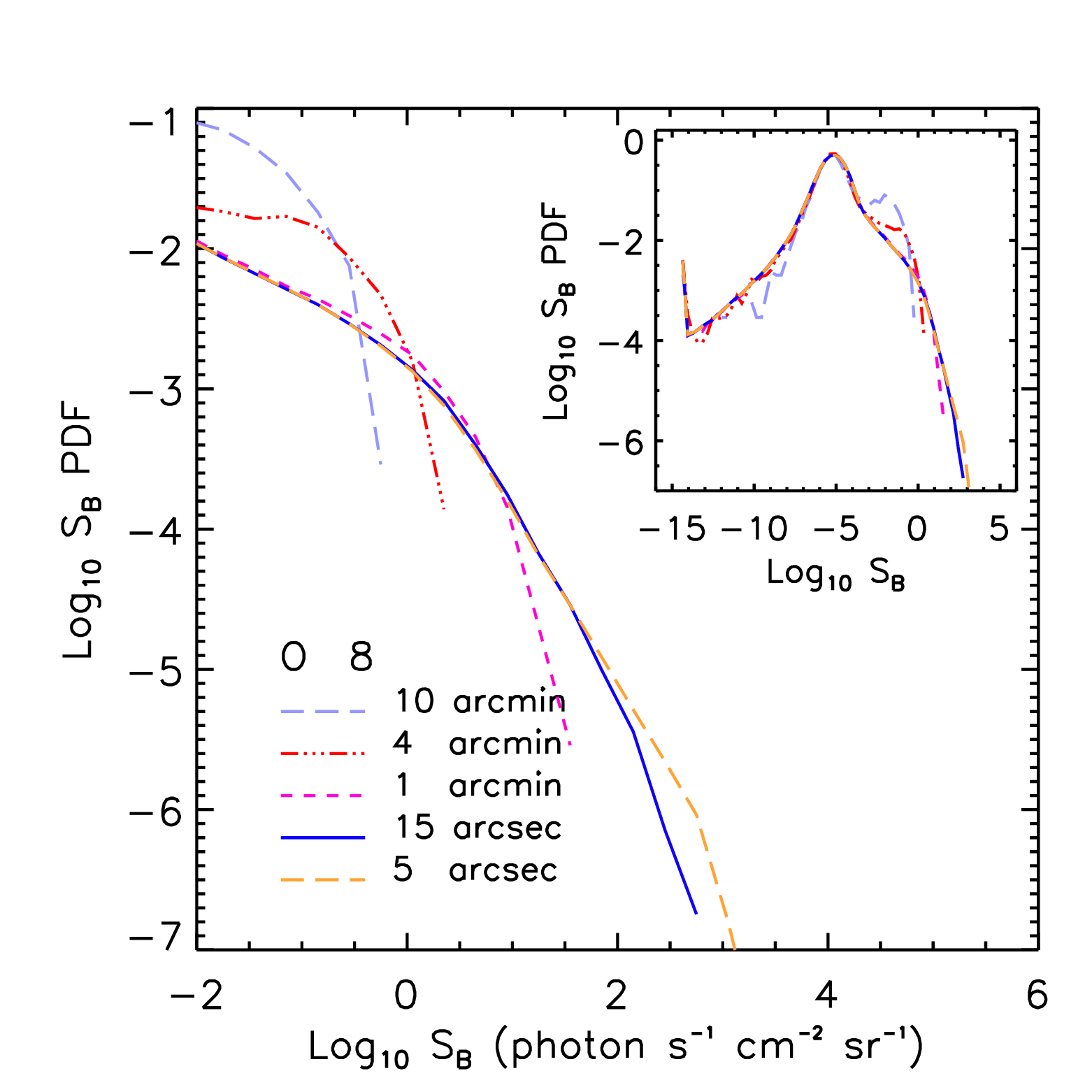}
\caption{As Fig.~\protect\ref{em_pdf}, but for \oviii\ emission as a
  function of 
  the angular resolution of the emission maps. The
  pixel sizes are $\vartheta = 5"$, $15"$, $1'$, $4'$, and $10'$, corresponding to
  physical sizes of 14 \hm\ kpc, 42 \hm\ kpc, 166 \hm\ kpc, 665
  \hm\ kpc and 1.661 \hm\ Mpc, respectively. Results
  broadly converge at high angular resolution ($\vartheta \lesssim
  1'$), except for the highest fluxes at the centres of haloes, which
  arise from gas in clouds with physical dimensions smaller than the
  pixel sizes considered here. There is no clear convergence at low
  angular resolutions ($\vartheta >1'$).}
\label{angle_figure}
\end{figure}

In this Section we investigate the effect of varying the angular
resolution of the detector.

Fig.~\ref{angle_figure} compares the PDFs of the measured flux of
\oviii\ emission assuming angular resolutions of
$\vartheta = 5"$, $15"$, $1'$, $4'$ and $10'$. The distributions
roughly converge for low 
fluxes, but there is no convergence at the highest fluxes and reducing
the pixel size continues to increase the highest observed flux, at
least down to the smallest scale considered here. This is because larger pixel
sizes prevent the detection of very strong emission coming from
structures with an angular size smaller than that used to observe
them.  When observing diffuse gas on scales larger then galactic
scales (such as the WHIM), an angular resolution of about 10-15" is
sufficient at $z\sim 0.25$ and does not imply a significant loss of
information. However, higher angular resolution might be desirable
when observing gas with more clumpy spatial distribution on smaller
scales, such as outflowing gas in galactic winds.  For reference, in
the \owls\ cosmology a pixel size of 15" subtends a comoving (proper)
size of about 52 (42) \hm\ kpc at $z=0.25$, 98 (66) \hm\ kpc at
$z=0.5$ and 175 (87) \hm\ kpc at $z=1$.

\section{Evolution}
\label{redshift}

\begin{figure}
\centering \includegraphics[height=8.4cm]{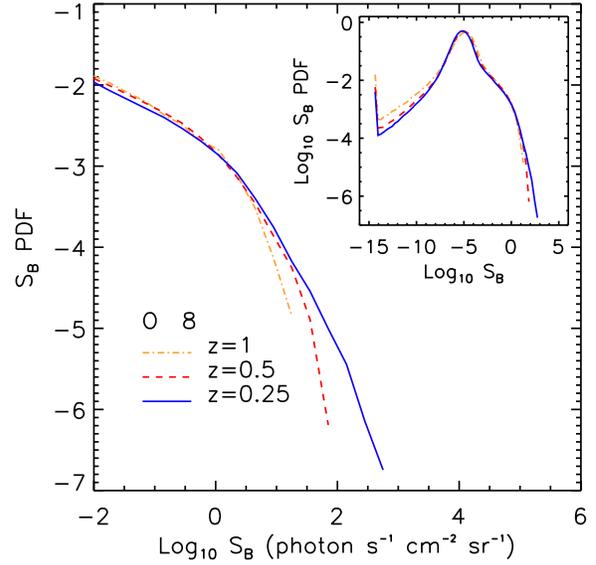}
\caption{As Fig.~\protect\ref{em_pdf}, but for \oviii\ emission as a function of
  redshift. Results are shown at $z=0.25$, 0.5 and 1. Maps are created
  assuming constant angular resolution $\vartheta = 15"$, which
  implies the physical size of a pixel increases with
  redshift from 42~\hm\ kpc at $z=0.25$ to 88~\hm\ kpc at $z=1$. The
  flux PDFs are nearly the same, but diverge 
  at the highest fluxes. In particular, the highest fluxes are found
  at the lowest redshift. This is probably mostly due to the varying
  physical size of the pixels.} 
\label{redshift_pdf}
\end{figure}

In this Section we briefly discuss how the intensity of the emitted
flux varies with cosmic time.

As mentioned in the previous Section, as redshift increases, the
physical size that corresponds to a pixel of a constant angular size
increases steadily, at least up to $z=1$. Moreover, the intrinsic
emission depends on the temperature, density 
and metallicity of the gas itself, as well as on the
UV/\xray\ background, all of which evolve. In Fig.~\ref{redshift_pdf}
we show the flux PDF of \oviii\ emission at 
redshifts $z=0.25$, 0.5 and 1.  The overall shape of the flux PDF
changes little with redshift, but the change is most noticeable at the
highest fluxes, where the maximum flux decreases with increasing
redshift. This is at least partly due to the fact that at constant
angular resolution and over the redshift range considered here, pixels
correspond to larger physical sizes at increasing redshifts and, as
shown in Section \ref{angle}, the maximum flux decreases with
increasing pixel size.

\section{What type of gas dominates the emission?}
\label{em_from}

In this Section we investigate what type of gas contributes most of the soft
\xray\ emission. We begin by showing the emission-weighted gas
distribution in the temperature-density plane for several lines in Section \ref{em_plane}. In Section~\ref{correlations} we then correlate the \oviii\ flux from individual particles  with their density, temperature and metallicity. Finally, we calculate the \oviii\ emission from gas in different 
density and temperature ranges in Section~\ref{cuts}.

\subsection{Temperature-density plane}
\label{em_plane}

\begin{figure}
\hspace{-0.9cm}
\includegraphics[width=9.5cm]{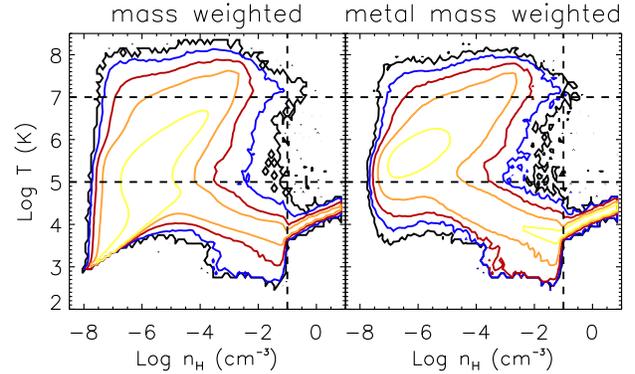}
\caption{The mass-weighted (left panel) and metal
  mass-weighted (right panel) temperature-density distribution of
  gas at $z=0.25$. The large
  difference between the two distributions implies that the bulk of
  the metals do not trace the bulk of the gas mass.
  The contours are logarithmically spaced by 1.5 dex. The
  horizontal lines at constant temperature highlight the WHIM range,
  while the vertical line at $n_{\rm H}=0.1$ cm\3\ indicates the limit
  above which we impose an equation of state onto star-forming particles. These
  dense particles are excluded from the calculation of the emission. For reference, the cosmic mean density corresponds to $n_{\rm H}\approx 3.7\times 10^{-7}\,\cm^{-3}$.}
\label{massdis}
\end{figure}

\begin{figure*}
\hspace{-1cm}
\includegraphics[height=18.5cm,angle=90]{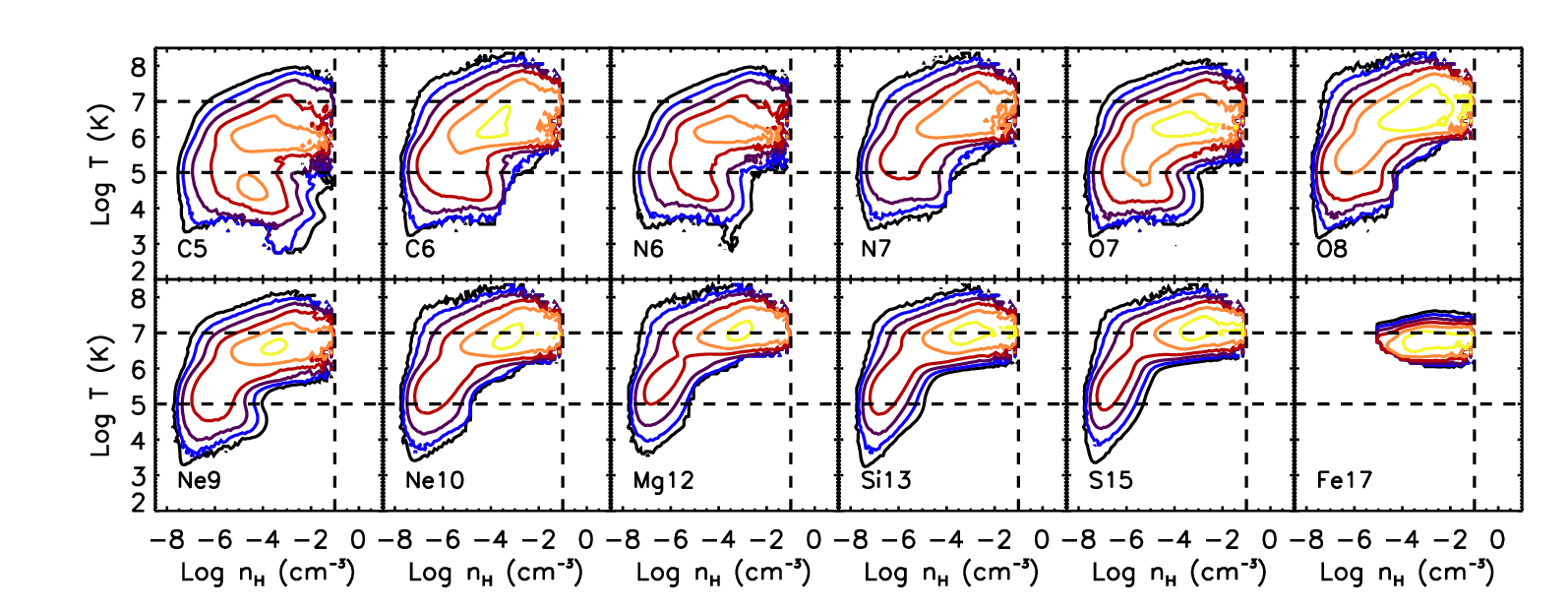}
\caption{The emission-weighted temperature-density distributions of
  gas at
  $z=0.25$. The distributions are weighted by the
  emission in different line for different panels, as indicated in
  the bottom-left corner of each 
  panel. From left to right and from top to bottom, the lines are:
  \cv, \cvi, \nvi, \nvii, \ovii, \oviii, \neix, \nex, \mgxii, \sixiii,
  \sxv\ and \fexvii. The contours are logarithmically spaced by 1.5
  dex and represent equal emission levels in all panels. The
  horizontal dashed lines at constant temperature highlight the WHIM
  range, while the vertical line at $n_{\rm H}=0.1$ cm\3\ indicates
  the limit above which we impose an effective equation of state onto
  star-forming gas. We do not include emission from this high-density
  gas. For reference, the cosmic mean density corresponds to $n_{\rm H}\approx 3.7\times 10^{-7}\,\cm^{-3}$. Lines from atoms with low atomic numbers are better tracers
  of WHIM emisison, while lines from heavier elements and from
  hydrogen-like ions mostly trace hotter gas with temperatures in
  excess of $10^7\,\K$. \fexvii\ emission is only present in a very narrow
  temperature range, while \cvi\ and \oviii\ lines trace gas in a wide
  temperature range.}
\label{xcont}
\end{figure*}

Fig.~\ref{massdis} compares the mass-weighted and the
metal mass-weighted temperature-density distributions of gas
at $z=0.25$. The vertical line at $n_{\rm
  H}=0.1$ cm\3\ indicates the density threshold above which
star-forming particles are put on a power-law equation of state 
(see Section~\ref{owls}) and the horizontal lines at $T=10^5\,\K$ and $T=10^7$
K delimit the temperature range that defines the WHIM.

The mass-weighted and the metal mass-weighted distributions in
Fig.~\ref{massdis} clearly indicate that the bulk of the metals do not
trace the bulk of the gas mass. A large fraction of the gas mass
resides in the low-density IGM with $T<10^5\,\K$, while metals are more
common in the moderately dense WHIM (see also
\citealt{wiersma2009b}). Also, a significant fraction of metals are
found in the relatively dense gas which populates the branch of the
distribution with $n_{\rm H} > 10^{-2}\,\cm^{-3}$ and $T<10^5\,\K$. This gas is
accreting onto galaxies and may eventually end up in the interstellar
medium (ISM).  

Fig.~\ref{xcont} presents the emission-weighted temperature-density
distributions of the gas, where the weight is given by the
emission in a different line in each panel. The lines we consider here
are \cv, \cvi, \nvi, \nvii, \ovii, \oviii, \neix, \nex, \mgxii,
\sixiii, \sxv\ and \fexvii, as listed in Table~\ref{eltable}.
Comparison of the mass-weighted, metal mass-weighted, and the 
emission-weighted temperature-density distributions demonstrate that
neither the bulk of the mass nor the bulk of the metals are good
tracers of the bulk of the soft \xray\ emission from metal lines. The peak of the
emission-weighted distributions of all lines is shifted towards higher
densities, compared to the mass- and metal mass-weighted
distributions. In fact, the emission of virtually all highly ionised
ions emitting in the \xray\ band concentrates in a region of the
temperature-density plane with $n_{\rm H} > 10^{-5}$ cm\3\ and $T\ga
10^6\,\K$. While the emission from hydrogen-like atoms is strong also
for temperatures in excess of $10^7\,\K$, emission from helium-like
atoms such as \cv, \nvi\ and \ovii\ peaks at $T < 10^7\,\K$ and declines
steeply at higher temperatures.

Fig.~\ref{xcont} demonstrate that, among the ones shown here, the best
lines to trace the WHIM in the temperature range $10^6\,\K<T<10^7\,\K$
are \cvi, \nvi, \nvii, \ovii, \oviii\ and \neix. All other ions tend
to trace hotter and somewhat denser gas with $T\ga 10^7\,\K$ and $n_{\rm
  H} > 10^{-4}\,\cm^{-3}$.

\subsection{Correlations between luminosity and gas properties}
\label{correlations}

\begin{figure*}
\centering
\includegraphics[width=\textwidth]{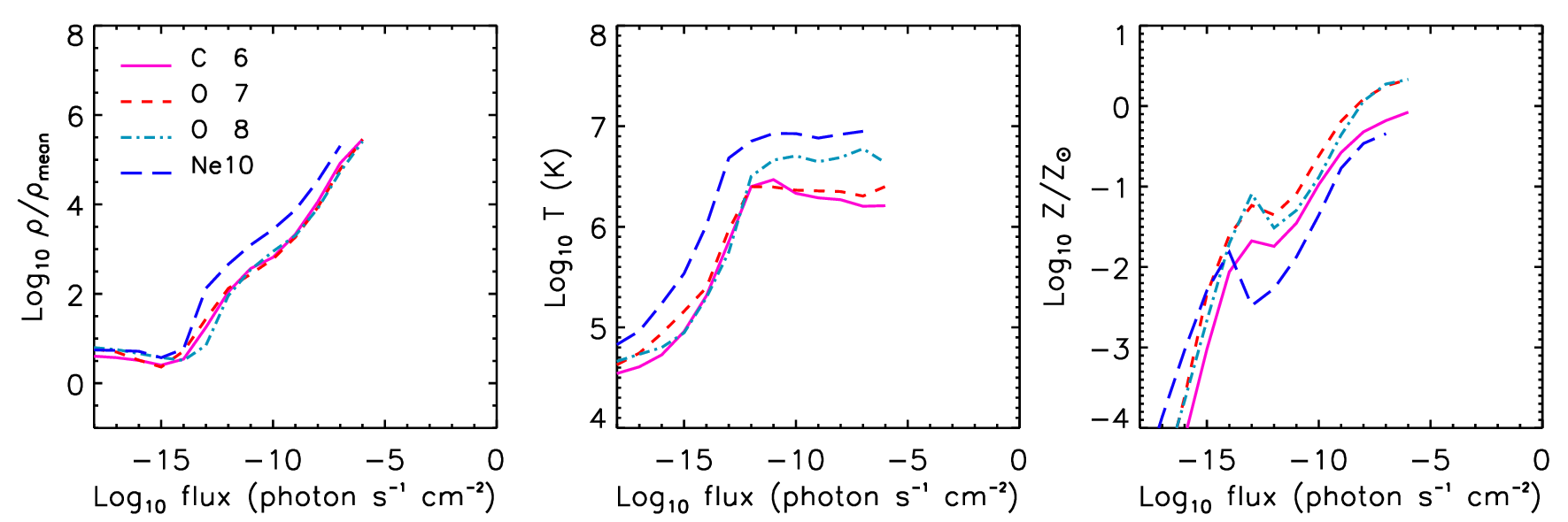}
\caption{Median values of the gas particles's overdensity (left panel),
  temperature (middle panel) and metallicity (right panel) that
  produce a given \xray\ particle flux at $z=0.25$ for the lines \cvi,
  \ovii, \oviii\ and \nex. Note that fluxes are given in units of
  photon s$^{-1}$ cm$^{-2}$, but that previous figures used \phot. Because we plot the flux emitted per particle, it is directly dependent on the numerical resolution. The absolute normalization of the $x$ therefore does not contain any useful information. For
  reference, $\rho_{\rm mean}(z=0.25)$ corresponds to $n_{\rm
    H}\approx 4\times 10^{-7}\,\cm^{-3}$ and our star formation
  threshold corresponds to $\log_{10} \rho/\rho_{\rm mean} \approx
  5.4$. Note that emission from higher density gas (i.e.\ the ISM) was
  ignored. The median density and metallicity   
  increase with luminosity, while the median temperature flattens out
  at the peak temperature of the emissivity of each line.}
\label{correl}
\end{figure*}

Fig.~\ref{xcont} shows what densities and temperatures dominate
the global emissivity in various emission lines, but much of this
emission may come out at fluxes that are too faint to be
detectable. It is therefore also of interest to investigate physical
conditions in the gas that produces the brightest emission.

\begin{figure*}
\centering
\includegraphics[height=\textwidth,angle=90]{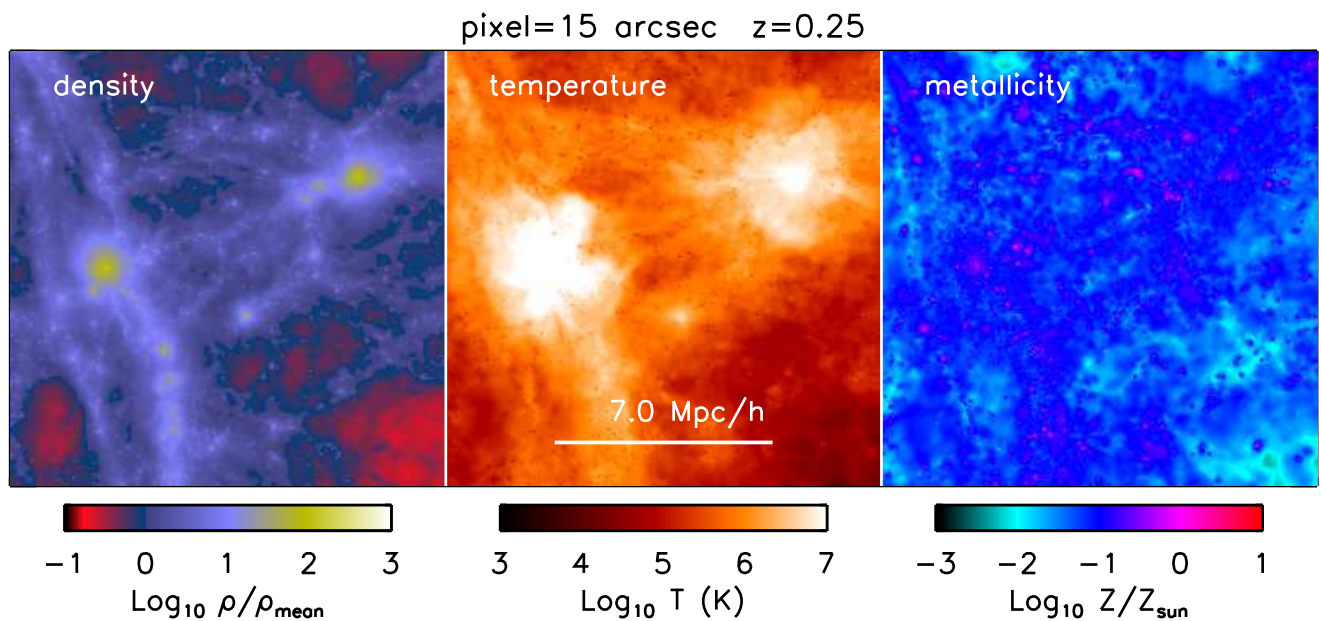}
\caption{As Fig.~\protect\ref{em_maps}, but showing maps of the gas overdensity (left panel), temperature (middle panel) and metallicity (right panels).}
\label{dtz}
\end{figure*}

Fig.~\ref{correl} investigates quantitatively the dependence of \cvi, \ovii, \oviii\ and \nex\ emission on the gas properties for a sample of emission lines by showing the median values of the particle density (left
panel), temperature (middle panel) and metallicity (right panel) that
produce a given particle luminosity. Since we consider individual
particles for this plot, the density, temperature and metallicity in
Fig.~\ref{correl} are local quantities, and as such cannot be compared
directly with those shown in Fig. \ref{dtz} in the following Section
and with the average quantities 
inferred observationally for the ICM and IGM. This is important for the
metallicity, as the metal distribution in our simulations is
inhomogeneous on small scales.
The scatter of the distributions, which is not shown, is relatively
small at high fluxes and increases at lower fluxes to up to 0.5 dex. 

The trends observed in Fig.~\ref{correl} are consistent with the
results of Fig.~\ref{xcont}. The line flux is strongly correlated with
density and metallicity, and the highest fluxes are always produced by
dense and metal-rich gas near galaxies. These correlations are
not surprising, because the emission scales with the element
abundance and with the density squared.  On the other hand, the median
gas temperature that produces a given flux is almost constant at the
highest fluxes and quickly decreases at the lowest fluxes. This is a
direct consequence of the fact that the line emissivity peaks in a
narrow temperature range, as shown in Fig.~\ref{lines}. The specific
value of the median temperature at the highest fluxes is almost
equivalent to the temperature where the line emissivity peaks, as also
seen in Fig.~\ref{xcont}. This suggests that the detection of a
specific emission line yields the temperature of the gas with a small
degree of uncertainty. The value of the observed flux then provides
information about the density and the metallicity of the gas. The
degeneracy between the two might be hard to break without further 
data from multiple lines from different elements and ionisation
states, but it appears from Fig.~\ref{correl} that, at the bright end,
the flux increases faster with the gas density than with the
metallicity. 

For each line, the maximum flux is reached at the density that
corresponds to our star formation threshold. Because we lack both the
physics and the resolution to model the gas at higher densities
(recall that we impose an effective equation of state onto the ISM),
we have ignored emission from higher densities. The peak fluxes are
therefore limited by this density cut. We stress that this means that
the models presented here underestimate the maximum fluxes. On the
other hand, the higher fluxes arise in gas with densities typical of
the ISM and can therefore hardly be counted as WHIM.

Finally, we note that the trends with density and temperature remain
the same if we assume a constant metallicity rather than the metal
distribution predicted by the simulation.

\subsection{Density and temperature cuts}
\label{cuts}

\begin{figure*}
\centering
\includegraphics[height=\textwidth,angle=90]{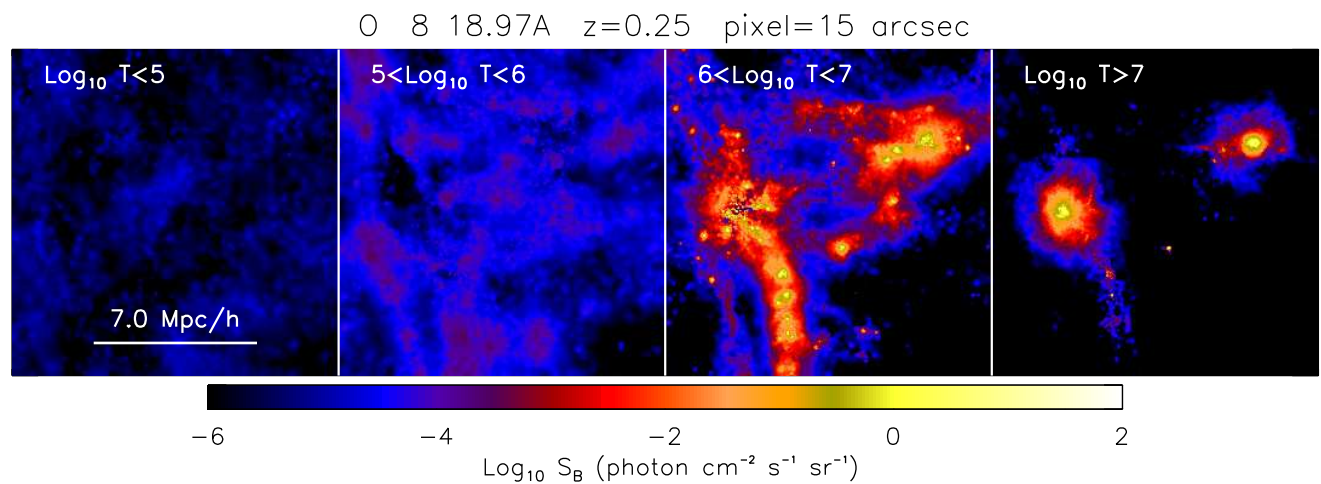}
\caption{As Fig.~\protect\ref{em_maps}, but showing maps of
  \oviii\ emission from gas in four different 
  temperature ranges. From left to right: $T<10^5\,\K$, $10^5\,\K$
  $<T<10^6\,\K$, $10^6\,\K$ $<T<10^7\,\K$ and $T>10^7\,\K$. The highest
  \oviii\ fluxes are produced by gas with $T>10^6\,\K$. \oviii\ emission
  is strongest in the highest temperature range.}
\label{tempcut}
\end{figure*}

\begin{figure*}
\centering
\includegraphics[height=\textwidth, angle=90]{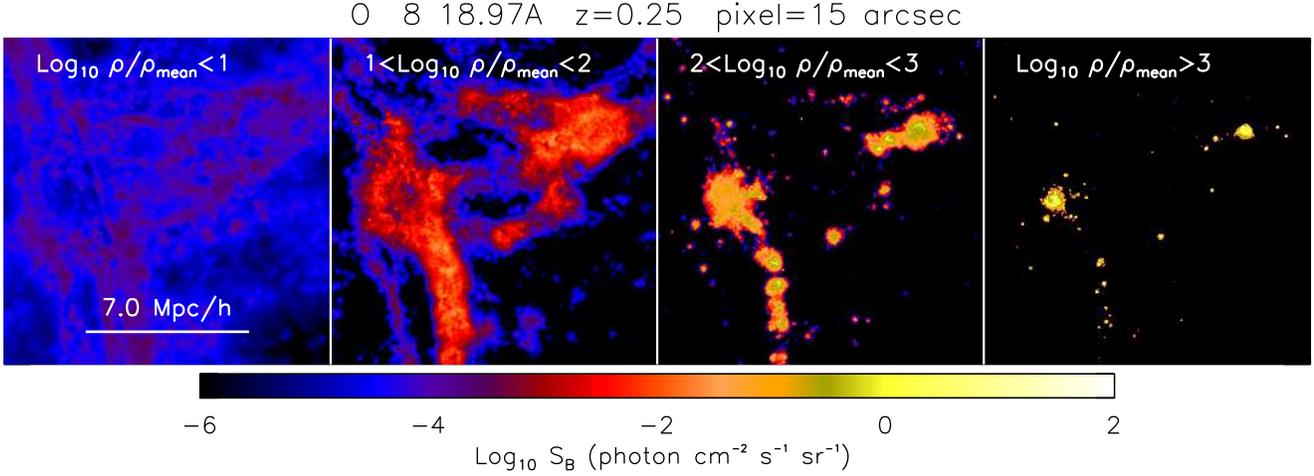}
\caption{As Fig.~\protect\ref{em_maps}, but showing maps of
  \oviii\ emission for gas in four different density 
  ranges, as indicated at the top of each panel. The highest fluxes are
  associated with the densest gas.}
\label{dencut}
\end{figure*}

\begin{figure}
\centering \includegraphics[width=8.4cm]{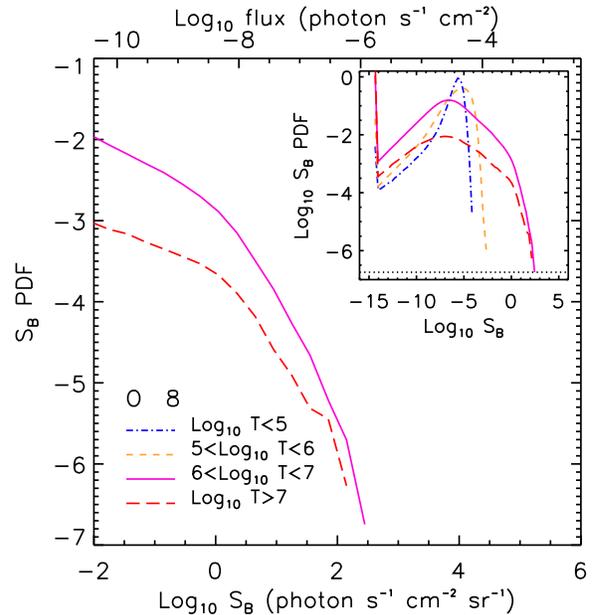}
\caption{As Fig.~\protect\ref{em_pdf}, but showing the PDF of the
  \oviii\ flux from gas in different temperature
  intervals.  The highest
  \oviii\ flux is produced by gas with $T>10^6\,\K$. \oviii\ emission is
  weakest in the lowest temperature ranges.}
\label{temp_pdf}
\end{figure}

\begin{figure}
\centering \includegraphics[width=8.4cm]{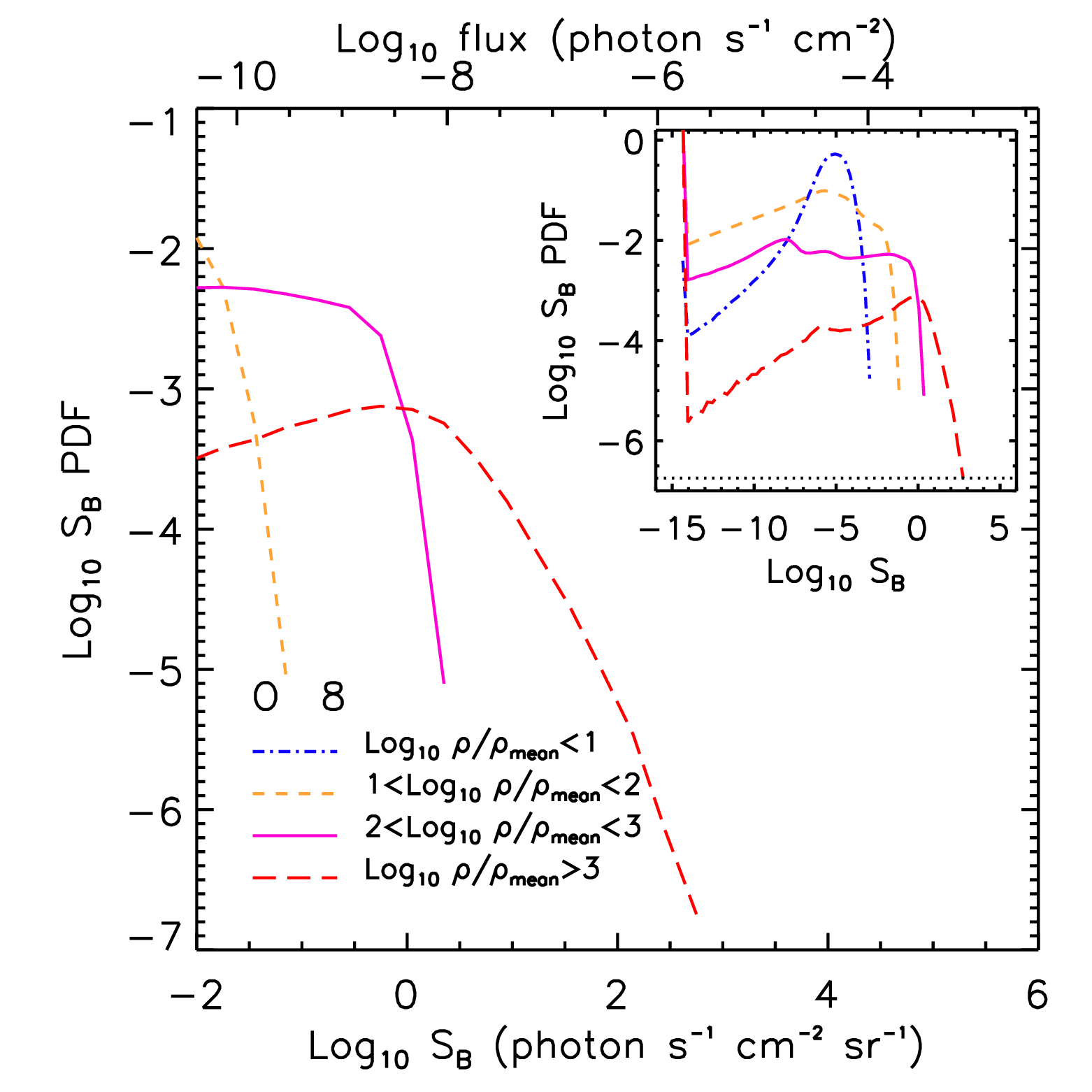}
\caption{As Fig.~\protect\ref{em_pdf}, but showing the PDF of the
  \oviii\ flux from gas in different density
  intervals. The highest
  fluxes are associated with the densest gas. In particular, the
  flux predicted for gas outside the virial radii of 
  haloes (i.e.\ $\rho / \rho_{\rm mean} \lesssim 200$) is at least a few
  orders of magnitude weaker than the flux from denser regions.}
\label{dense_pdf}
\end{figure}

In this Section we create \oviii\ emission maps using only gas
with temperatures and densities in a 
given interval. This will help us to understand the dependence of the
emission on temperature and density and will give an indication of
what type of gas produces what kind of emission.  For comparison,
we show in Fig.~\ref{dtz} the density, temperature and metallicity of the
gas in the same region of space shown in all the emission maps in
this paper.

Fig.~\ref{tempcut} shows \oviii\ emission maps made using only gas
with temperatures in a specific range. We consider the
four temperature intervals $T<10^5\,\K$, $10^5\,\K<T<10^6\,\K$, $10^6\,\K
<T<10^7\,\K$ and $T>10^7\,\K$. Similarly, in Fig.~\ref{dencut} we
consider four different density intervals, with $\textrm{Log}_{10}
\rho / \rho_{\rm mean}$ varying by one dex in each density bin. The
flux PDFs for the entire simulation volume are shown in
Fig.~\ref{temp_pdf} for the temperature cuts 
and in Fig.~\ref{dense_pdf} for the density cuts.
Figs. \ref{tempcut}--\ref{dense_pdf} demonstrate that the strongest
\oviii\ emission comes from gas in dense ($\textrm{Log}_{10} \rho /
\rho_{\rm mean} > 3$) and hot ($T>10^6\,\K$) regions.

\section{Dependence on the physical models}
\label{em_model}

A key feature of the \owls\ project is the
availability of a large number of runs with varying physical
prescriptions. In this Section we investigate how a series of physical
prescriptions affects the predictions for the WHIM emission. We will
only show results for \oviii\ $\lambda 18.97$ \AA\ emission, but note
that we find similar trends for other lines. We focus on
a subset of the \owls\ models with varying feedback schemes and
cooling recipes, which are the physical implementations that produce
the largest variations in the diffuse emission. These runs differ not
only in the total amount of metals produced and in the spatial
distribution of the metals, but also in the star formation history
\citep{schaye2010} and
in the density and temperature distribution of the diffuse gas.  
Physical prescriptions are changed one at a time, in order to isolate
their impact on the results.  The physical models considered here are
the following 
(we refer the reader to \citet{schaye2010} and to the works cited
below for more detailed descriptions):

\begin{enumerate}

\item \default: Our reference model, as described in
  Section~\ref{owls}. All results presented in the preceding sections
  were obtained from this simulation. This model uses the star
  formation prescription 
  of \citet{schaye2008}, the supernova-driven wind model of
  \citet{vecchia2008} with wind mass loading $\eta = 2$ and wind
  initial velocity $v_{\rm w}=600~\kms$, the metal-dependent cooling
  rates of \citet{wiersma2009a}, the chemical evolution model
  described in \citet{wiersma2009b} and the WMAP3 cosmology. AGN
  feedback is not included.

\item \nosn: As \default, but without supernova
  feedback. The metals produced by stars are only transported by gas
  mixing and no energy is transferred to the IGM and ISM when SNe
  explode.

\item \zcool: As \default, but with radiative cooling
  rates calculated assuming primordial abundances. The
  neglect of the contributions of metal lines implies slower gas
  cooling than in the \default\ model.

\item \agn: As \default, but with the addition of AGN
  feedback as described in \citet{booth2009};

\item \wml: As \default, but with a wind mass
  loading twice as high, that is $\eta = 4$. The total energy injected
  in winds is thus also twice that in \default;

\item \wdens: As \default, but the wind properties scale with the local
  gas density as (\citealt{schaye2010}):
\begin{eqnarray}
v_{\rm w} &=& v_{\textrm{w}0} \left(\rho / \rho_{\rm crit} \right)^{1/6}\\
\eta &=& \eta_0 \left( \rho / \rho_{\rm crit}\right)^{-2/6}
\end{eqnarray}
where $\rho_{\rm crit}$ is the density threshold for star
formation, $v_{\textrm{w}0}= 600~\kms$ and
$\eta_0 = 2$. This implies that at $\rho_{\rm crit}$ the gas is kicked
with the same velocity and mass-loading as in the \default\ model.
This particular scaling of wind velocity with density was chosen so
that the wind velocity scales with the local sound speed  (recall that
we impose the equation of state $P\propto \rho^{4/3}$ onto the ISM) and the
wind mass-loading is scaled such that the total wind energy is
independent of the gas density. As a consequence, the wind energy is
equal to that in the \default\ model.

\item \wmom: As \default, but with ``momentum--driven'' galactic winds
  following the prescription of \citet{oppenheimer2008}, with the
  difference that winds are ``local'' to the star formation event and
  are fully hydrodynamically coupled as in the \default\ model (see
  \citealt{vecchia2008} for a discussion of the importance of these
  effects).  The 
  wind initial velocity and the wind mass loading factor are defined
  as a function of the galaxy velocity dispersion $\sigma$ as $v_{\rm
    w} = 5\sigma$ and $\eta = v_{\textrm{w}0}/\sigma$ respectively,
  with $v_{\textrm{w}0}=150~\kms$. The velocity dispersion $\sigma =
  \sqrt{2}v_{\rm circ}$ is estimated using an on-the-fly
  friends-of-friends halo finder. This model was motivated by the idea
  that galactic winds may be driven by radiation pressure on dust grains
  rather than supernovae. The energy injected into the wind
  becomes much higher than for \default\ for halo masses exceeding $10^{11}
  - 10^{12}\,{\rm M}_\odot$.

\item \mill: As \default, but using the WMAP year 1
  cosmology \citep{spergel2003} as in the Millennium simulation
  \citep{springel2005} and a wind mass loading factor $\eta =
  4$. Results from this model must be compared to those of the
  \wml\ model to isolate the effect of cosmology.

\begin{figure*}
\centering
\includegraphics[width=0.8\textwidth]{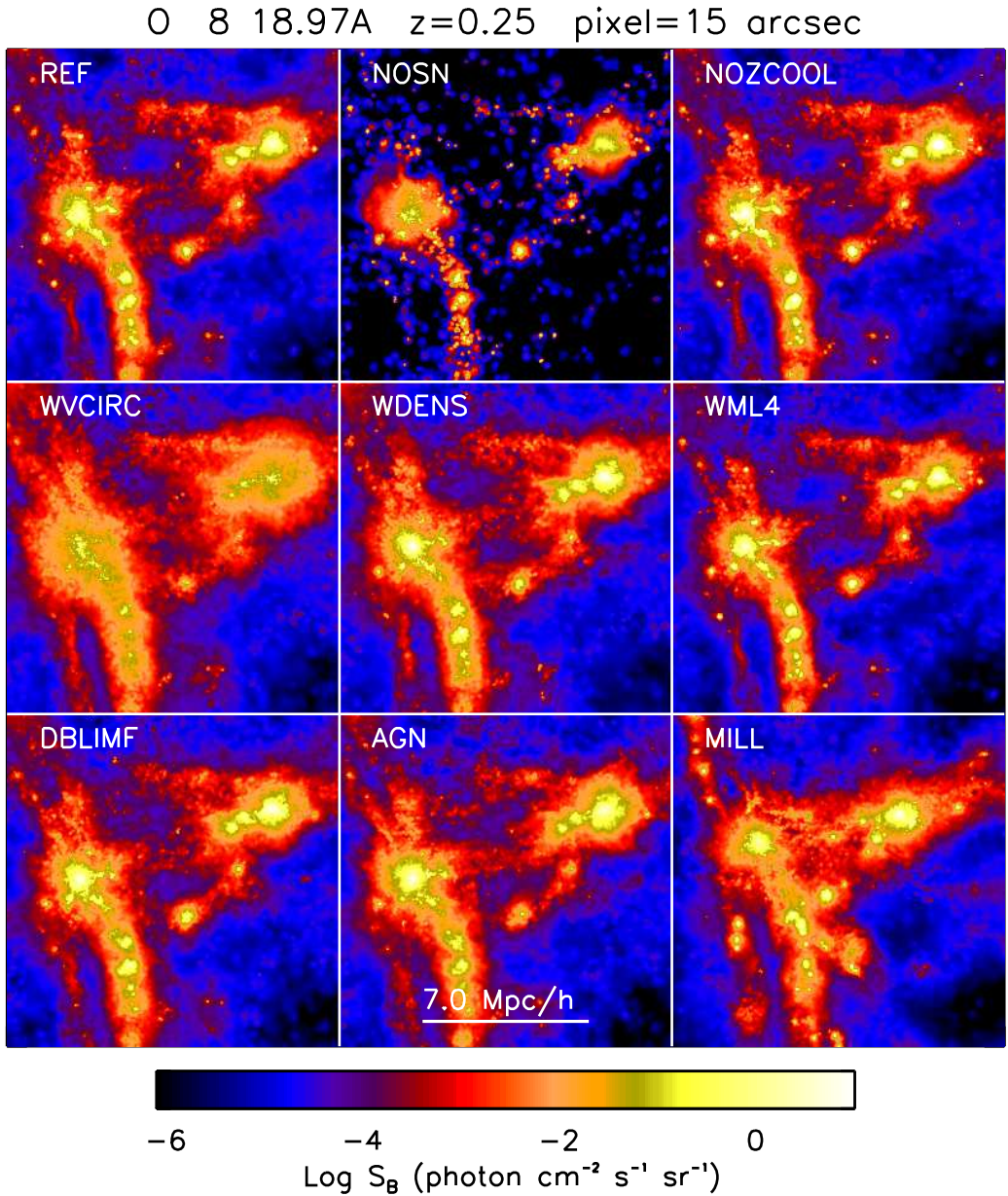}
\caption{As Fig.~\protect\ref{em_maps}, but showing \oviii\ emission
  maps for different simulations, as indicated in the top left
  corner of each panel.}
\label{sims}
\end{figure*}

\item \dblimf: As \default, but using a top-heavy IMF at high gas
  pressures. This model switches from a Chabrier IMF to a power-law
  IMF $dN/dM\propto M^{-1}$ (as compared to 
  $\propto M^{-2.3}$ for the high-mass tail of the Chabrier IMF) at
  the critical
  pressure $P/k=2.0\times 10^6\,\cm^{-3}\,\K$, which was chosen
  because for this value $\sim 10^{-1}$ of the stellar mass forms at
  higher pressures. The extra SN energy is used to increase the
  initial wind velocity to $1618~\kms$ for star particles formed with a
  top-heavy IMF, while keeping the mass loading factor fixed at
  $\eta=2$ (as for the Chabrier IMF, these parameter values
  correspond to 40\% of the available SN energy). This model is called
  \emph{DBLIMFCONTSFV1618} in the \owls\ project (which includes
  more models with top-heavy IMFs in starbursts, see
  \citealt{schaye2010}). 
\end{enumerate}

\begin{figure*}
\centering
\includegraphics[width=0.8\textwidth]{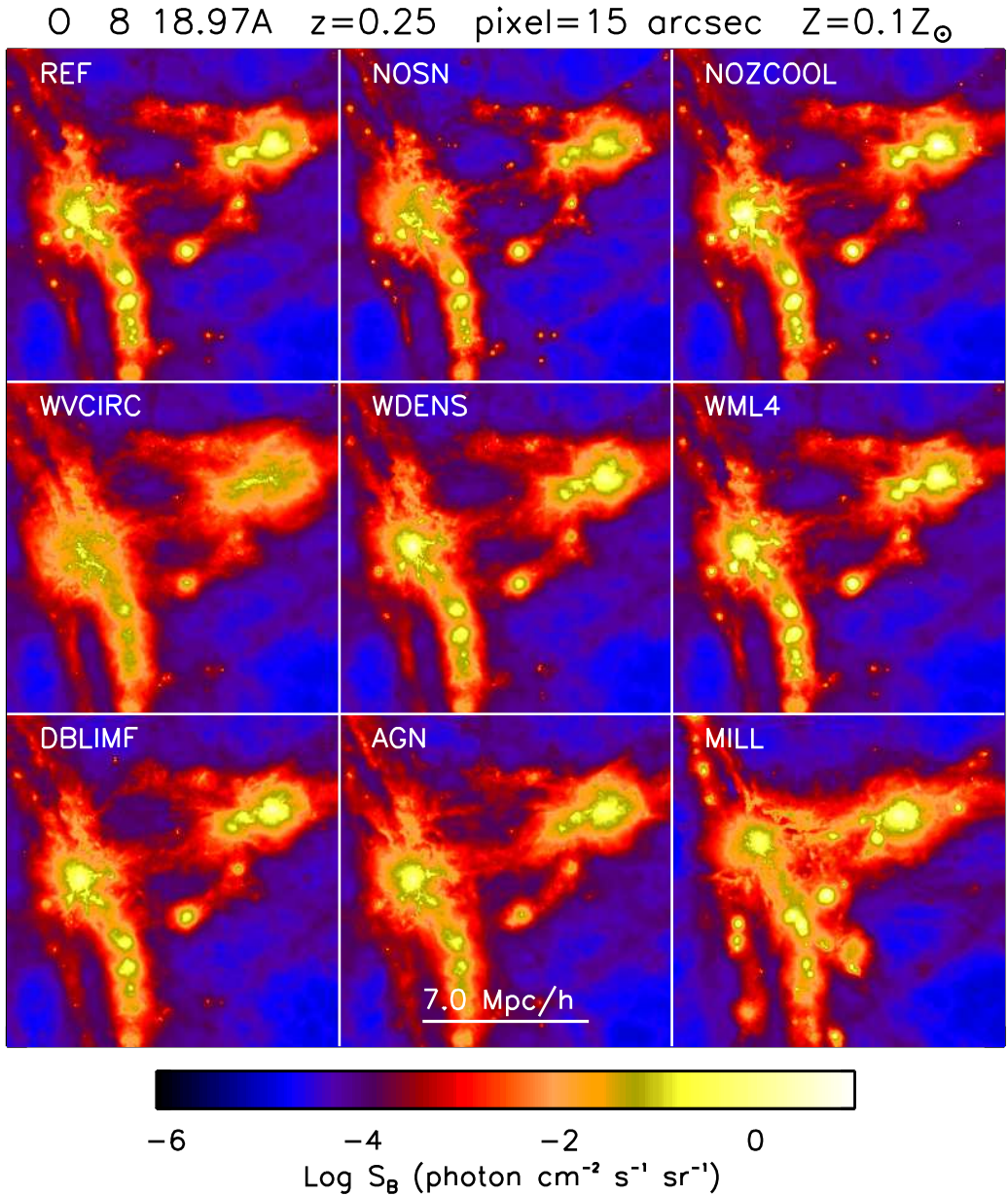}
\caption{Same as Fig. \ref{sims}, but for constant gas metallicity $Z=0.1 Z_{\sun}$.}
\label{simszconst}
\end{figure*}

\begin{figure*}
\centering
\includegraphics[width=8.7cm]{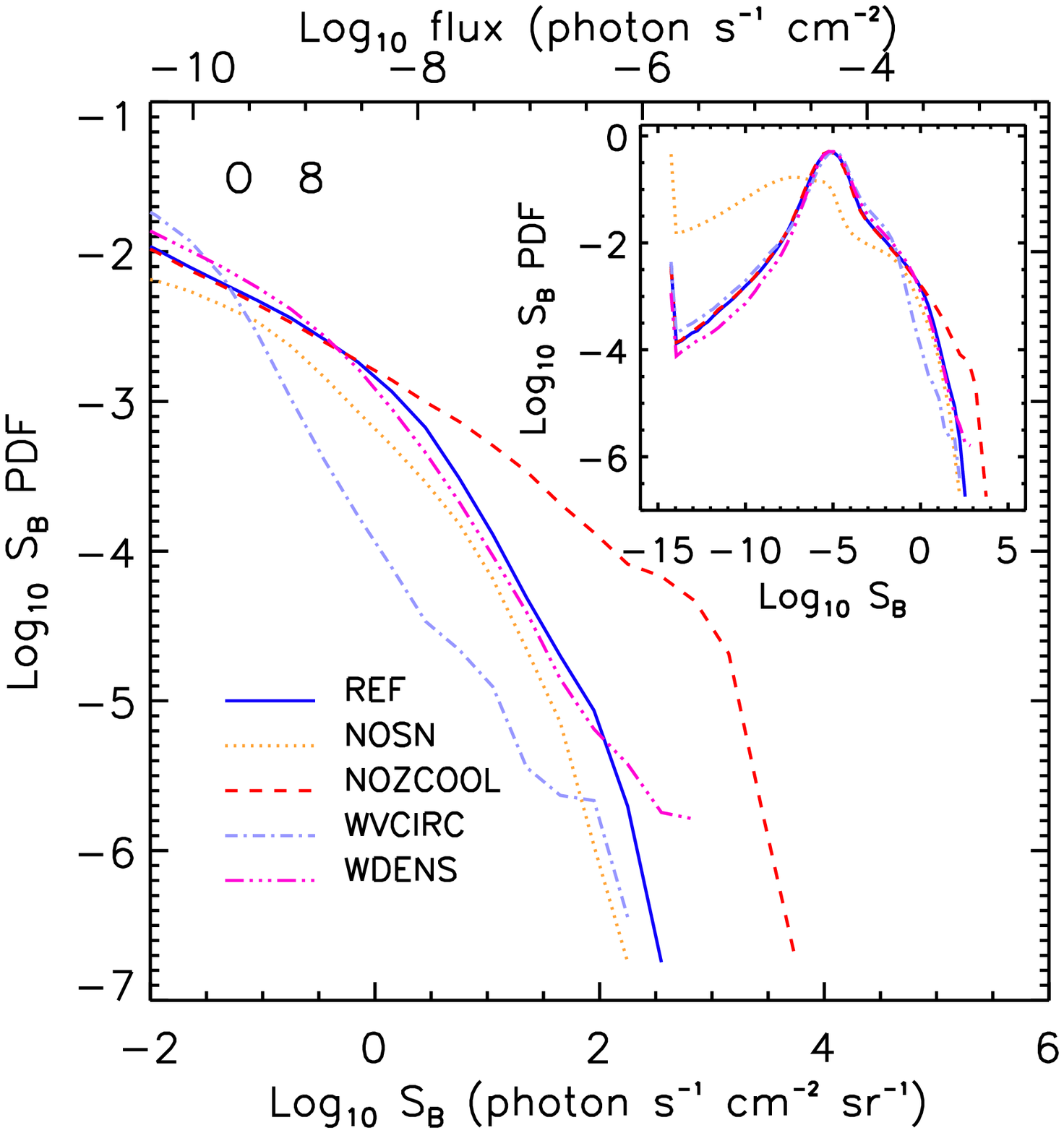}
\includegraphics[width=8.7cm]{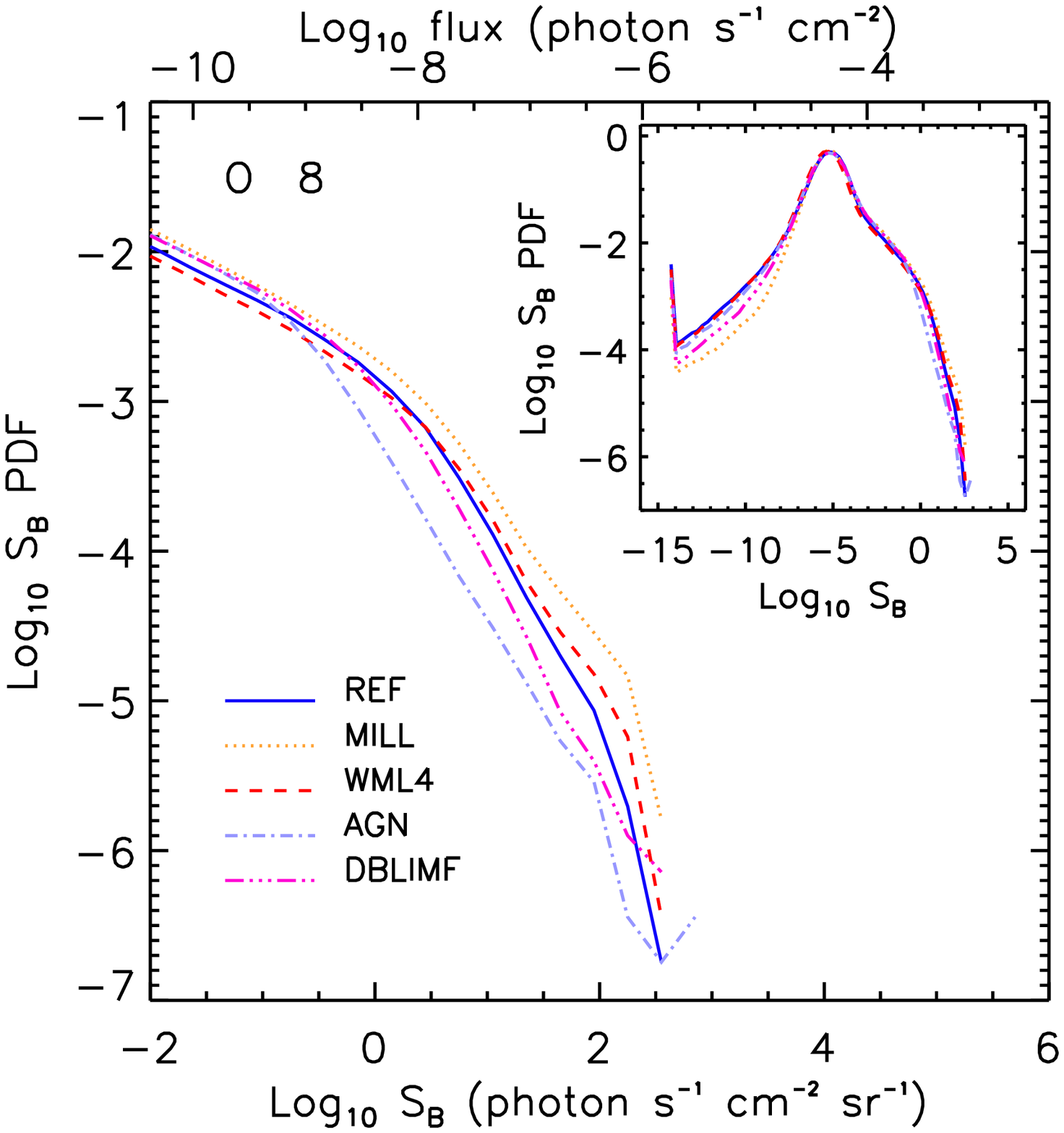}
\caption{As Fig.~\protect\ref{em_pdf}, but showing the flux PDFs of
  \oviii\ emission for simulations 
  with varying physical prescriptions. The simulations shown in the
  left panel are the same ones as in Fig.~\ref{sims}. Additional runs
  are shown in the right panel and compared to the results of the
  \default\ run. The results are surprisingly robust to changes in the
physical model. However, neglecting metal-line cooling results in a
large overestimate of the maximum flux. Ignoring supernova-driven
winds has little effect for the brightest pixels, but otherwise shifts
the PDF to lower fluxes.} 
\label{sims_pdf}
\end{figure*}

\begin{figure*}
\centering
\includegraphics[width=8.7cm]{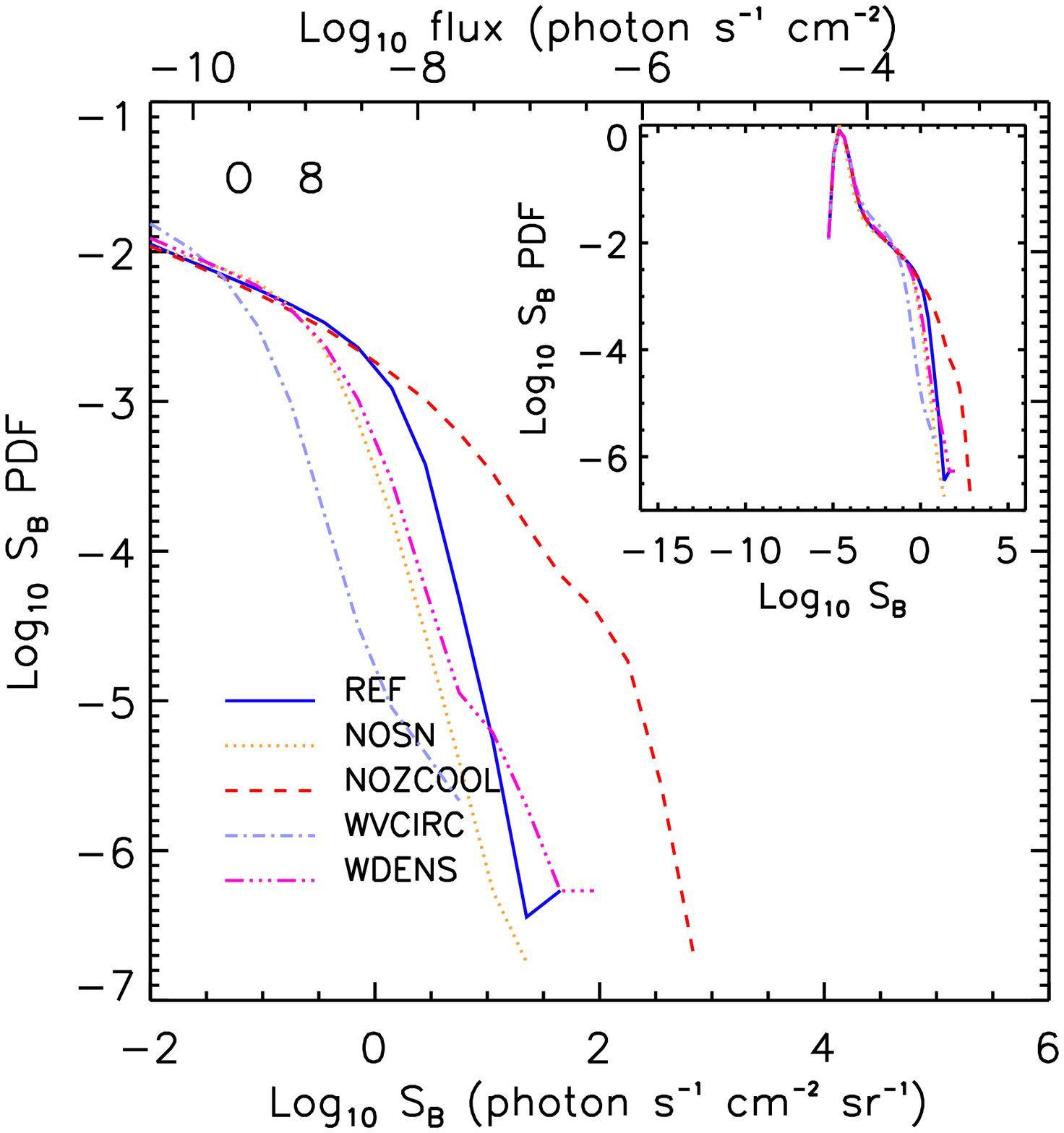}
\includegraphics[width=8.7cm]{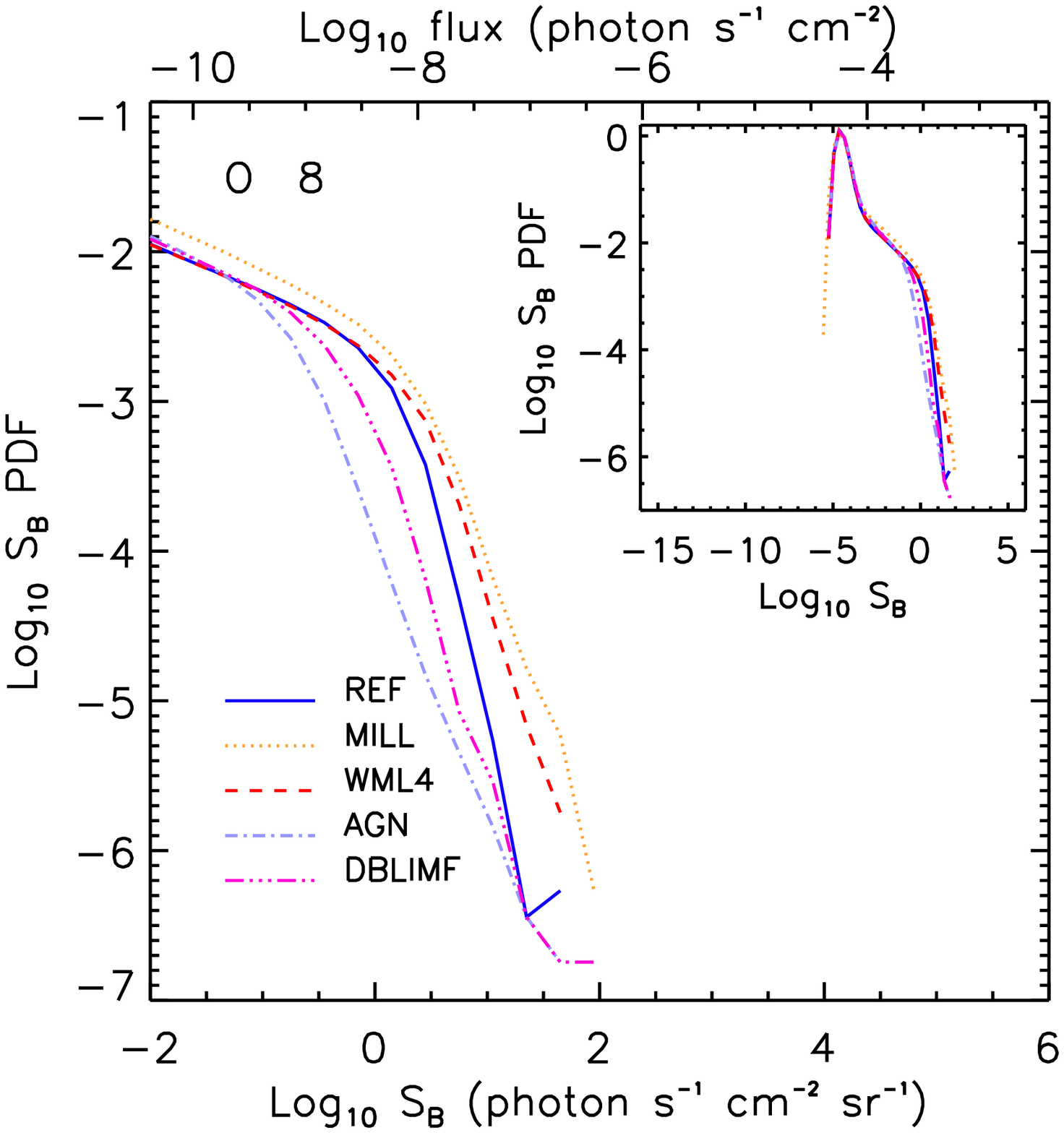}
\caption{As Fig.~\protect\ref{em_pdf}, but showing the flux PDFs of
  \oviii\ emission for simulations 
  with varying physical prescriptions, under the assumption of
  a constant gas metallicity of $Z=0.1 Z_{\sun}$. The simulations shown
  are the same ones as in Fig.~\ref{sims}. The
  assumption of constant metallicity boosts the flux from the
  low-density, cool regions of the universe, where the largest 
  contribution to the line emissivity comes from photo-ionisation by
  the UV background (compare the inset to that of
  Fig.~\ref{sims_pdf}), but it reduces the maximum fluxes. The
  differences between the models are qualitatively the same as when
  the predicted metal distributions are used (compare with
  Fig.~\ref{sims_pdf}).} 
\label{zconst_sims_pdf}
\end{figure*}

Fig.~\ref{sims} shows maps of the \oviii\ emission for all
runs in a region of 14 \hm\ Mpc on a side or, equivalently, 1.12
degrees on the sky. Fig. \ref{simszconst} shows the emission in the same simulations, but for a constant gas metallicity of $Z=0.1 Z_{\sun}$.
Figs.~\ref{sims_pdf} and \ref{zconst_sims_pdf} present the flux 
PDFs of the \oviii\ emission using the predicted abundances and
$Z=0.1 Z_{\sun}$, respectively.

The dominant impression we get from looking at these figures, is that
most models predict very similar \oviii\ fluxes. This is somewhat
surprising, as the physics that is included is quite different and the
models predict a rather wide range of star formation histories
\citep{schaye2010}. Closer inspection does, however, reveal several
differences. Visually, we can easily see from the top panels of
Fig.~\ref{sims} that models \nosn, \wmom, and \mill\ differ most
from our reference model.
Comparing Figs.~\ref{sims} and \ref{simszconst}, we see that for \nosn\ the
difference is driven by the metal distribution. Without galactic
winds, the metals remain mostly confined to the inner parts of haloes and
the flux from low-density gas is sharply reduced. The
predicted metal distributions can, however, not account for the
differences seen for \wmom, and \mill. The flux from high-density gas
is considerably lower for \wmom, which indicates that the peak
densities are lower and/or that the temperatures of high-density gas
are lower than the value for which the emissivity of \oviii\ peaks
(i.e.\ $10^{6.3}\,\K$). The \mill\ run appears different mostly
because structure formation has progressed further in this run,
because it uses a higher value of $\sigma_8$.

The flux PDFs shown in Figs.~\ref{sims_pdf} and \ref{zconst_sims_pdf}
confirm the conclusions drawn from the visual inspection of the
emission maps. Looking first at the global distributions shown in the
insets, we see that only \nosn\ differs substantially, the peak of the
PDF having shifted to much lower fluxes. If a constant
metallicity is assumed, then all models agree near the peak of the
distribution and the low-flux tail of the PDF vanishes. This indicates
that fluxes $\ll 10^{-5}\,$\phot\ imply gas metallicities
$\ll 0.1 Z_\odot$. For all models the peak fluxes are lower if a
constant metallicity is assumed, from which we conclude that the high
flux tail probes gas with metallicities in excess of $0.1 Z_\odot$. 

The model for which the high-flux tail differs the most is \zcool,
which neglects metal-line cooling. Its
flux PDF starts to differ from that of \default\ for fluxes $\ga
1\,$\phot, reaching a maximum flux that is about an order of magnitude
higher. The fact that metal-line
cooling is important is not surprising, as the metal-line emission that
we are studying dominates the radiative cooling rates for temperatures
typical of the WHIM and for metallicities $\ga 10^{-1}\,Z_\odot$
\citep[e.g.][]{wiersma2009a}. Ignoring metal-line cooling during the
simulations leaves the gas in the temperature-density regime where it
emits efficiently, artificially boosting the predicted fluxes.
The calculation of the emission from metals in this
model is of course inconsistent with the assumption of primordial abundances
used to calculate the gas cooling rates. However, this inconsistency
is present in other work \citep[e.g.][]{fang2005}, which often used
cooling rates based on primordial abundances during the simulations
and calculates the metal-line emission of the gas in
post-processing. Clearly, this approach will strongly overestimate the
maximum fluxes.
 
While the emission map of the \mill\ model appeared to be
substantially different (Fig.~\ref{sims}), the change in cosmology
does not change the statistics of the \oviii\ emission substantially, as
can be seen by comparing the flux PDFs of the \wml\ and
\mill\ models. 

There are enormous differences in the
feedback processes that are included in the simulations shown here,
ranging from no galactic winds (\nosn), to efficient winds driven by
supernovae (\wdens), to extremely strong winds in starbursts resulting
from a top-heavy IMF (\dblimf), to violent AGN feedback (\agn). These
differences strongly affect the star formation histories
\citep{schaye2010}, yet the flux PDFs of all these models are
strikingly similar. This 
suggests that the metal distributions are also similar in the gas that
emits \oviii\ efficiently. Indeed, we will show elsewhere (Wiersma et
al., in preparation; McCarthy et al., in preparation) that the
metallicity of the WHIM (much more so than that of other components)
is remarkably robust to changes in the feedback prescription. Indeed,
assuming a constant metallicity does not change the relative
difference between the models much. This suggests that the differences
in the high flux tails are driven by differences in the distribution
of the baryons in the temperature ($T\sim 10^{6.5}\,\K$) and density
($\rho \gg 10^3\rho_{\rm mean}$) regime that dominates the emission
from the brightest regions (see Fig.~\ref{correl}). In particular, it
appears that the ``momentum-driven'' winds (\wmom), which predict
substantially lower peak fluxes (even for a constant metallicity),
substantially reduce the peak gas densities of the WHIM. This suggests that
these winds are able to remove more gas from the centers of haloes, perhaps
because the winds in this model use much more energy than is available
from supernovae. 

Thus, we conclude that bright \oviii\ emission, and other soft X-ray lines too, may tell us more about the gas densities in the centers of haloes than about the metallicities and the temperatures (which must be close to the peak of the emissivity curve for the emission to be bright). This is because the
emissivity scales as $Z\rho^2$ and is thus most sensitive to the gas density.

\section{Can the WHIM be detected in emission?}
\label{detect}

\begin{table*}
\centering
\caption{Summary of the technical characteristics of the instruments on \ixo\ and \xenia.}
\begin{tabular}{l c c c c c}
\hline
\hline
Instrument & Effective area & FoV & Angular resolution & Energy range & Energy resolution \\
 & (m$^2$) & (arcmin$^2$) & (arcsec) & (keV) & (eV) \\
\hline
\ixo\ & & & & & \\
XMS  & 3 @ 1.25 keV & $5.4 \times 5.4$ & 5 & $0.3-7.0$ & 2.5 @ central $2'\times 2'$ \\
WFI  & 3 @ 1.25 keV & $18 \times 18$ & 5 & $0.1-15$ & 150 \\
\hline
\xenia\ & & & & & \\
WFS & 0.12 @ 0.6 keV & $42 \times 42$ & 222 & $0.1-2.2$ & 3 @ 0.5 keV \\
WFI & 0.058 @ 1 keV & $60 \times 60$ & 15  & $0.3-6$ & 70 @ 1 keV \\
\hline
\hline
\end{tabular}
\label{exper}
\end{table*}

In this Section we discuss the possibility that current and future
\xray\ telescopes may detect the WHIM emission at the levels predicted
by our simulations.

Previous studies predict that \xray\ emission from the WHIM is far below the detection threshold of the \chandra\ and \xmm\ telescopes (\citealt{Yoshikawa2003}; \citealt{Yoshikawa2004}; \citealt{fang2005}). Predictions that WHIM emission lines in filaments could be detected by \xmm\ have not been confirmed \citep{pierre2000}. However, future missions such as 
\edge\ \citep{piro2009}, \ixo\ and \genx\ \citep{windhorst2006}, have been
shown to have a better chance of detecting diffuse emission (\citealt{Yoshikawa2004}; \citealt{fang2005}; \citealt{bregman2009}).

Ideally, the emission from the WHIM requires an instrument
which couples a large field of view (FoV, hereafter) with high angular
and spectral resolution (\citealt{Yoshikawa2004}; \citealt{fang2005};
\citealt{paerels2008}). While instruments with high angular and
spectral resolutions alone are fine for searches of absorption features, a large FoV is desirable for mapping diffuse emission. High angular resolution is also very important, as we found that the emission is dominated by bright, compact regions. Without sufficient angular resolution, it will not only be more difficult to detect the emission, but one would also risk misinterpreting its origin. For example, emission that looks smooth and filamentary at poor angular resolution may come from unresolved, dense knots rather than from diffuse, low density gas.

A few dedicated small to medium-size missions have been proposed to investigate the WHIM emission. The Missing Baryon Explorer (\mbe, \citealt{fang2005}) was the first of such medium-size missions, later followed by the Diffuse Intergalactic Oxygen Surveyor (\dios, \citealt{ohashi2006}), the Extreme phySics in the TRansient and Evolving cosMOs (\estremo, \citealt{piro2006}), \edge\ \citep{piro2009} and \xenia\ \citep{hartmann2009}. The last two missions are very similar. The common goal of all these missions is to build an instrument with a large FoV and high spatial and spectral resolution.

The proposed technical specifications of the instruments we consider are listed in Table \ref{exper}.
The FoV of the \xray\ Microcalorimeter Spectrometer\footnote{http://ixo.gsfc.nasa.gov/technology/xms.html} (XMS) on board \ixo\ is $5.4' \times 5.4'$, which corresponds to 1.1 \hm\ comoving Mpc at $z=0.25$. Also on \ixo, the Wide Field Imager (WFI, \citealt{treis2009}) has a FoV of $18' \times 18'$, corresponding to 3.7 \hm\ comoving Mpc at the same redshift. Both instruments have sufficient coverage to image gas in filaments at $z\sim 0.25$, with an angular resolution of 5" each.
For comparison, the Wide Field Imager (WFI) and the Wide Field Spectrometer (WFS) on \xenia\ were proposed to have FoVs of 1 and 0.7~degrees and angular resolutions of 15" and 3.7', respectively. While XMS on \ixo\ and WFS on \xenia\ have impressive energy resolutions of about 3~eV, it is much worse for the WFIs (150~eV for \ixo\ and 70~eV for \xenia). 

The unresolved cosmic X-ray background has been measured to be $\approx 13 \pm 1$~\phot\ keV$^{-1}$ in the range 0.5 -- 1~keV (\citealt{hickox2009}) and is thought to be dominated by diffuse Galactic and local thermal-like emission. For the energy resolution of the XMS ($\Delta E = 2.5$~eV) this corresponds to about $3\times 10^{-2}$~\phot. 

The instrumental background for XMS is about $2\times 10^{-2}$~photon s$^{-1}$ keV$^{-1}$ per cm$^{-2}$ of the detector area (Kaastra 2010, private communication). For a pixel size of $300~\mu$m and 3~arcsec this corresponds to $8.5\times 10^{11}$~photon s$^{-1}$ sr$^{-1}$ keV$^{-1}$. For an effective area of $2\times 10^4~\cm^2$ and an energy resolution of 2.5~eV this implies a limiting surface brightness of $1.1\times 10^{-2}$~\phot, slightly below the extragalactic background. 
Thus, we expect observations to become background limited below $10^{-1}$~\phot. For an angular resolution of 15~arcsec and an effective area of $2\times 10^4~\cm^2$, this surface brightness would result in about 10 detected photons per resolution element in a 1~Ms exposure. Hence, with an instrument like XMS it would be possible to detect surface brightnesses $\ga 1$~\phot\ (at 15 arcsec resolution) with high confidence in reasonable exposure times. Fig.~\ref{dense_pdf} shows that this would enable the detection of \oviii\ from gas with $\rho \ga 10^2 \rho_{\rm mean}$ at $z=0.25$. 

For the proposed WFS for \xenia\ the instrumental background is expected to be even lower and the energy resolution similar to that for XMS on \ixo. The angular resolution is, however, much poorer (about 4') and the effective area is an order of magnitude smaller ($10^3\,\cm^2$). A 1~Ms exposure would in that case yield $\sim 10^2$ photons per resolution element for a surface brightness of $10^{-1}$~\phot, more than sufficient for a robust detection. Note that XMS on \ixo\ would detect an order of magnitude more photons when averaged over the same angular size (because its effective area is an order of magnitude larger). We predict from the equivalent version of Fig.~\ref{dense_pdf} for a pixel size of 4 arcmin (not shown) that this surface brightness limit is sufficient to detect some \oviii\ emission from gas with $\rho > 10^3 \rho_{\rm mean}$ at $z=0.25$. Note that while this emission would appear to be diffuse due to the low angular resolution, it would in fact come from unresolved, dense clouds. 

Compared with the WFS, the WFI for \xenia\ has much better angular resolution (15"), a smaller effective area (about $0.6\times 10^3\,\cm^2$) and a much worse energy resolution ($\Delta E = 70$ eV). Because the spectral resolution is more than an order of magnitude lower than for the WFS, the backgrounds will be higher by the same factor. For reasonable exposure times ($\sim 1$~Ms) we would still not be background limited. The poor spectral resolution would, however, make it much more challenging to interpret the observations, as the lines correspond to the differential Hubble flow across a comoving distance $\sim 250~h^{-1}\,$Mpc (for $z=0.25$). On the other hand, the interpretation would be greatly facilitated by the much improved angular resolution, although this will not necessarily help with the identification of the lines.

The WFI for \ixo\ has an energy resolution that is lower still, $\Delta E = 150$ eV, which may lead to severe projection effects and which would make it challenging to identify the lines. While WFI does not have a better angular resolution than XMS, it does have a much larger field of view (18~arcmin vs.\ 2~arcmin for the high energy resolution mode of XMS). 

According to our predictions, the WFS on \xenia\ should be able to detect strong emission from the WHIM only for gas with overdensities greater than a thousand, such as in the centres of groups. \ixo\ would be extremely exciting as it would be sufficiently sensitive to detect emission from \oviii, and probably a few other lines, from both the cores of small groups and from the outskirts of larger groups and clusters down to overdensities $\ga 10^2$, and with an angular resolution that is sufficient to resolve much of the intrinsic structure. 

Instruments with higher sensitivity and lower flux limits will be
required in the future to image the emission from the diffuse WHIM in low-density regions and in filaments where $\rho \ll 10^2 \rho_{\rm mean}$. \genx\ \citep{windhorst2006} might be one of those missions.
Since most of the baryons
contained in the WHIM reside at such low overdensities
(Fig.~\ref{massdis}), it seems unlikely that the bulk of the baryons
will be detected in soft X-ray emission any time soon. Absorption
experiments clearly have a better chance to accomplish this goal.

However, we have seen that detecting soft X-ray emission from
the gas in and around (groups of) galaxies is within reach of proposed
missions. This is extremely exciting, as such observations would
provide vital clues on gas infall, outflows and metal enrichment,
which are the key, poorly understood processes in models of the
formation of galaxies.

\section{Conclusions}
\label{summary}

Gravitational accretion shocks  and, to a lesser extent, galactic winds driven by supernovae and/or AGN, can shock-heat the diffuse intergalactic medium
to temperatures $\ga 10^5\,\K$. At low redshifts a large fraction of
the baryons are thought to reside in this warm-hot intergalactic medium (WHIM; $10^5\,\K<T<10^7\,\K$) and soft X-ray emission lines from metals offer a way to
detect them. 

Here, we used a subset of cosmological simulations from the
OverWhelmingly Large Simulations (\owls) project \citep{schaye2010} to
study the nature and detectability of this emission. These simulations
are among the largest dissipative runs ever performed, and use new
prescriptions for a range of baryonic processes, such as star
formation, galactic winds driven by massive
stars and supernovae, 
and feedback from AGN. Of particular relevance for us, is that they are
fully chemodynamical, tracking the delayed release of 11 different
elements by massive stars, supernovae of types II and Ia, and
asymptotic giant branch stars, and that radiative cooling is implemented
element-by-element in the presence of an evolving UV/X-ray
background. 

Our main conclusions can be summarised as follows:
\begin{enumerate}
\item The strongest emission in the soft \xray\ band comes from the
  \oviii\ line, whose emissivity peaks at a temperature of a few times
  $10^6\,\K$. A large number of other lines are, however, only
  slightly weaker.

\item Emission lines from metals provide strong constraints on the
  temperature of the emitting gas. Most of the emitted radiation and
  essentially all of the potentially detectable flux
  comes from gas with temperatures close to the peak of the emissivity
  curve for a collisionally ionised gas.
  Because of the quadratic dependence of the emissivity on density,
  the emission traces neither the bulk of the baryonic nor the bulk of
  the metal mass. Instead, the emission is dominated by metal-rich gas
  at very high overdensities in and around galaxies. 
Since the peak temperature of the emissivity increases with atomic number and ionisation state, lines from heavier or more highly ionised atoms trace hotter gas. As such, \cvi\ and \ovii\ trace gas with $T\sim 10^6\,\K$, \oviii, \neix\ and \fexvii\ gas with $10^6\,\K<T<10^7\,\K$, and \nex, \mgxii, \sixiii\ and \sxv\ gas with $T\sim 10^7$ K.

\item By comparing results from different \owls\ models, we
  have investigated the dependence of the results on a number of
  physical mechanisms. It is essential to accurately model the cooling
  rates. In particular, studies that neglected metal-line cooling
  are not self-consistent and will have overestimated the maximum
  fluxes by about an order of magnitude. Without galactic winds, the PDF of the flux
  shifts to much lower values. However, the high-flux tail, which will
  become detectable first, is surprisingly robust to strong variations
  in the strength of the winds, remaining similar even when
  galactic winds are turned off or when violent AGN feedback is
  included. While assuming a constant metallicity of $0.1 Z_\odot$
  reduces the width of the flux PDF (and thus suppresses the maximum
  fluxes), the relative differences between the models remain
  similar. This suggests that the different feedback processes mainly
  affect the metal-line emission by changing the gas densities in and
  around galaxies, as the highest fluxes always come from regions with
  temperatures close to the peak of the emissivity curve and with
  (local) metallicities $\ga Z_\odot$.

\item Soft \xray\ emission lines in the low-redshift Universe are
  likely to be detectable only for relatively 
  high-density regions, such as groups and the outskirts of
  clusters. Because dense, compact structures can easily dominate the
  emission even if they contain only a very small amount of mass, it
  is important for future observatories to have high angular
  resolution. Without sufficient angular resolution, the emission is
  harder to detect and one risks misinterpreting the origin of the
  detected emission. For example, emission that looks diffuse may in
  reality come from a number of compact gas clouds in and around
  galaxies that trace the large-scale structure. 

\item While soft X-ray line emission is too faint for it to be a promising route to close the baryon budget, it does offer the exciting 
possibility to image the gas accreting onto and flowing out of
galaxies. Currently proposed missions could provide
vital clues on some of the most poorly understood, but crucial
ingredients of models of the formation and evolution of galaxies. 
\end{enumerate}

\section*{Acknowledgments}

We would like to thank Jelle Kaastra for invaluable help with the  physics of \xray\ emission and for suggesting to look at the different \ovii\ components. 
We are grateful to Volker Springel and all other members of the \owls\ team for their help during the development of this project and to Tesla Jeltema for reading the manuscript. We also acknowledge useful discussions with many members of the \estremo, \edge\ and \xenia\ teams. 
The simulations presented here were run on Stella, the LOFAR BlueGene/L system in Groningen, on the Cosmology Machine at the Institute for Computational Cosmology in Durham as part of the Virgo Consortium research programme, and on Darwin in Cambridge. This work was sponsored by the Dutch National Computing Facilities Foundation (NCF) for the use of supercomputer facilities, with financial support from the Netherlands Organization for Scientific Research (NWO).  SB acknowledges support by NSF Grant AST-0507117 and by STFC. This work was supported by Marie Curie Excellence Grant MEXT-CT-2004-014112, by an NWO VIDI grant, and was partly carried out under the HPC-EUROPA project, with the support of the European Community - Research Infrastructure Action of the FP7.

\appendix

\section{Convergence tests}
\label{converge}

In this Appendix we present a number of tests we have performed to
verify how numerical parameters affect our results. In particular, we consider
the effect of resolution in Appendix~\ref{number}. Results for different box
sizes (at constant resolution) are shown in Appendix~\ref{box}. In
Appendix~\ref{thick} we investigate the effect of the 
slice thickness on the shape of the flux PDF.  All tests use the
\default\ physical model and results are shown only for the \oviii\ 18.97
\AA\ line.

\subsection{Resolution}
\label{number}

\begin{figure}
\centering \includegraphics[width=8.4cm]{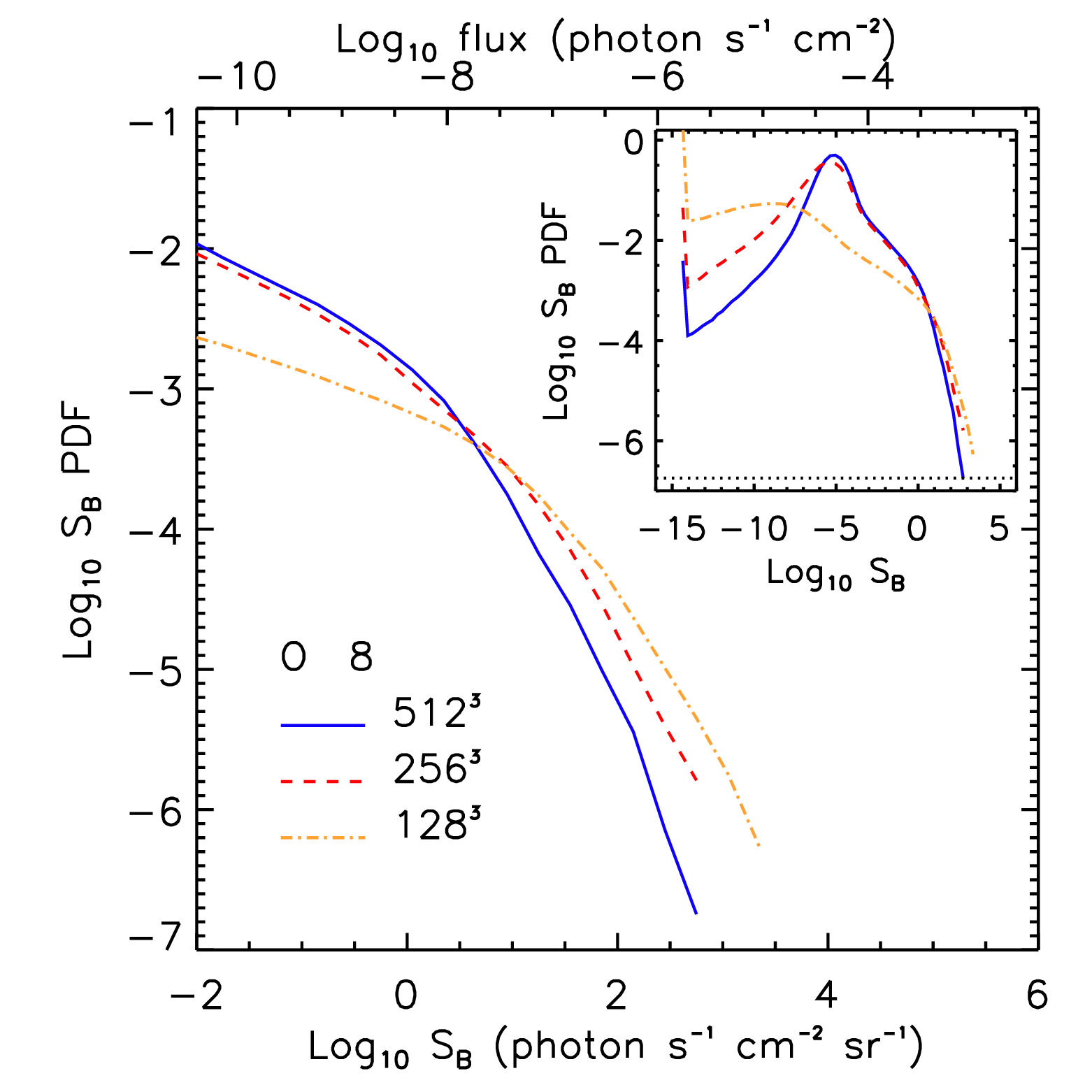}
\caption{As Fig.~\protect\ref{em_pdf}, but showing the \oviii\ flux
  PDF as a function of the 
  resolution of the simulation. The particle mass
  (gravitational softening) increases by a factor of eight (two) when the
  number of particles 
  decreases by a factor of $2^3$. All runs assume the same physical
  model, angular resolution and box size. The predictions of the
  $N=256^3$ and $512^3$ runs are very close for fluxes greater than
  $10^{-5}\,$\phot.} 
\label{number_figure}
\end{figure}

To test how the numerical resolution affects our results, we have
created \oviii\ emission maps for three different realisations of the
same simulations with particle numbers $N=2\times 128^3$, $2\times
256^3$ and the default number of $2\times 512^3$. The corresponding dark matter
particle masses are $2.6\times 10^{10}$, $3.2\times 10^9$ and $4.1\times
10^8\,$\hm\ \msun, respectively. The corresponding maximum physical
gravitational softening lengths are 8, 4, and 2\hm\ kpc. All runs use
a box size of 100 \hm\ comoving Mpc.

Fig.~\ref{number_figure} shows the flux PDF for maps with 15" angular
resolution. The lowest resolution run with $N=2\times 128^3$ deviates
most from the other two, in particular for fluxes lower than about 1
\phot. This is at least in part caused by the fact that this run
severely underestimates the 
stellar mass \citep{schaye2010} and hence the gas metallicity.
The higher resolution runs with $2\times 512^3$ and
$2\times 256^3$ particles show good agreement, although there are
small differences at the highest fluxes.

\subsection{Box size}
\label{box}

\begin{figure}
\centering \includegraphics[width=8.4cm]{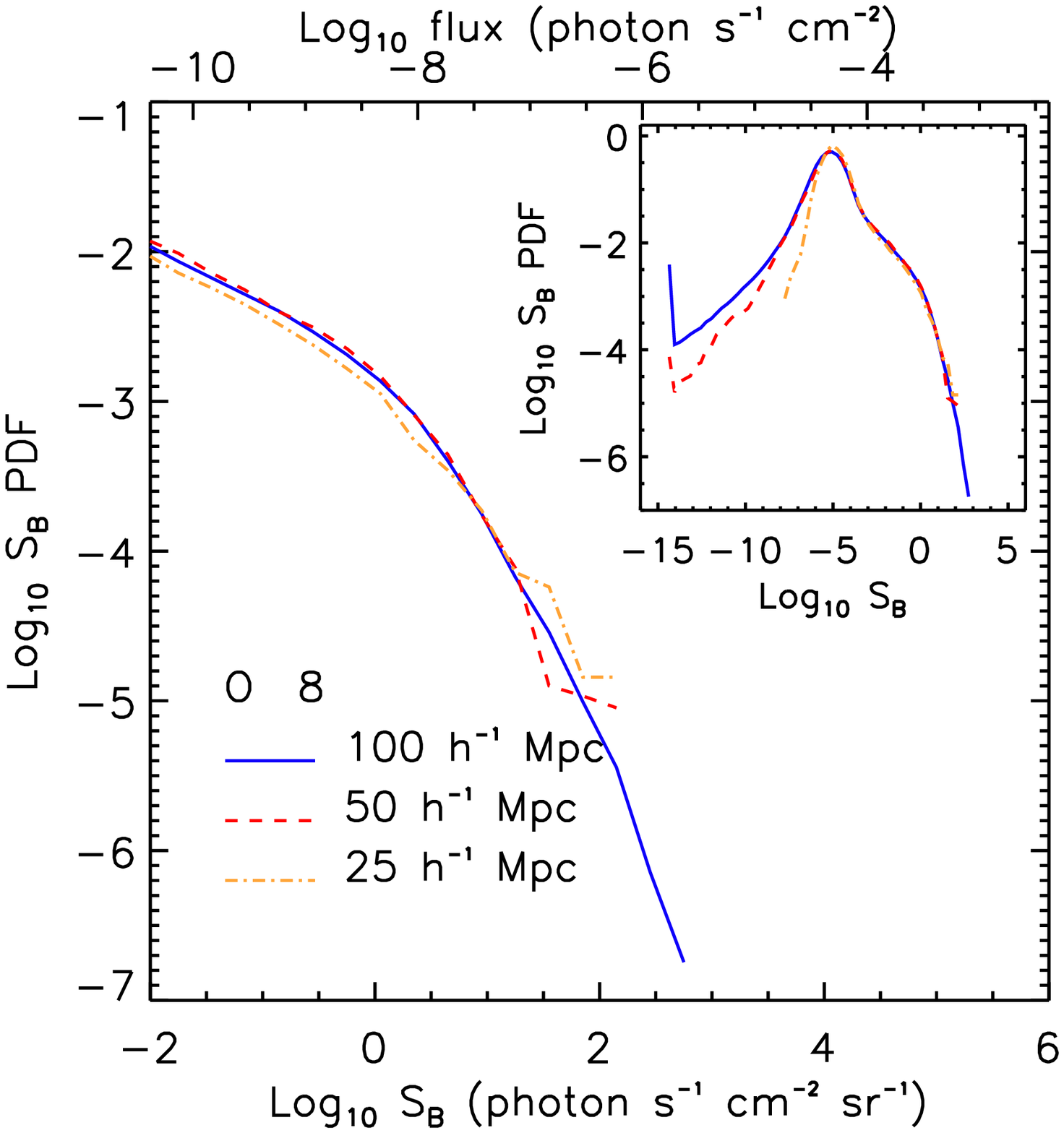}
\caption{As Fig.~\protect\ref{em_pdf}, but showing the \oviii\ flux
  PDF for simulations with
  box sizes of 100, 50 and 25 \hm\ Mpc. All
  runs assume the same physical model, particle mass and angular
  resolution. Except for very low fluxes ($<10^{-5}\,$\phot), the
  results are insensitive to the size of the simulation box.}
\label{box_figure}
\end{figure}

We have tested the effect of the size of the simulation box on the
flux PDF of \oviii\ emission, while keeping the resolution
constant. Fig.~\ref{box_figure} compares simulations implementing the
\default\ model with 
box sizes of 100, 50 and 25 \hm\ Mpc, respectively,
and with constant dark matter particle mass $m_{\rm DM} = 4.1\times 10^8$
\hm\ \msun. The
flux PDFs converge for intermediate flux values, but diverge at the
lowest fluxes. The lowest and very highest fluxes
are only found in the largest box, which contains the greatest number
of particles and provides the most pixels. It thus samples the tails
of the distribution best.

\subsection{Slice thickness}
\label{thick}
\begin{figure}
\centering \includegraphics[width=8.4cm]{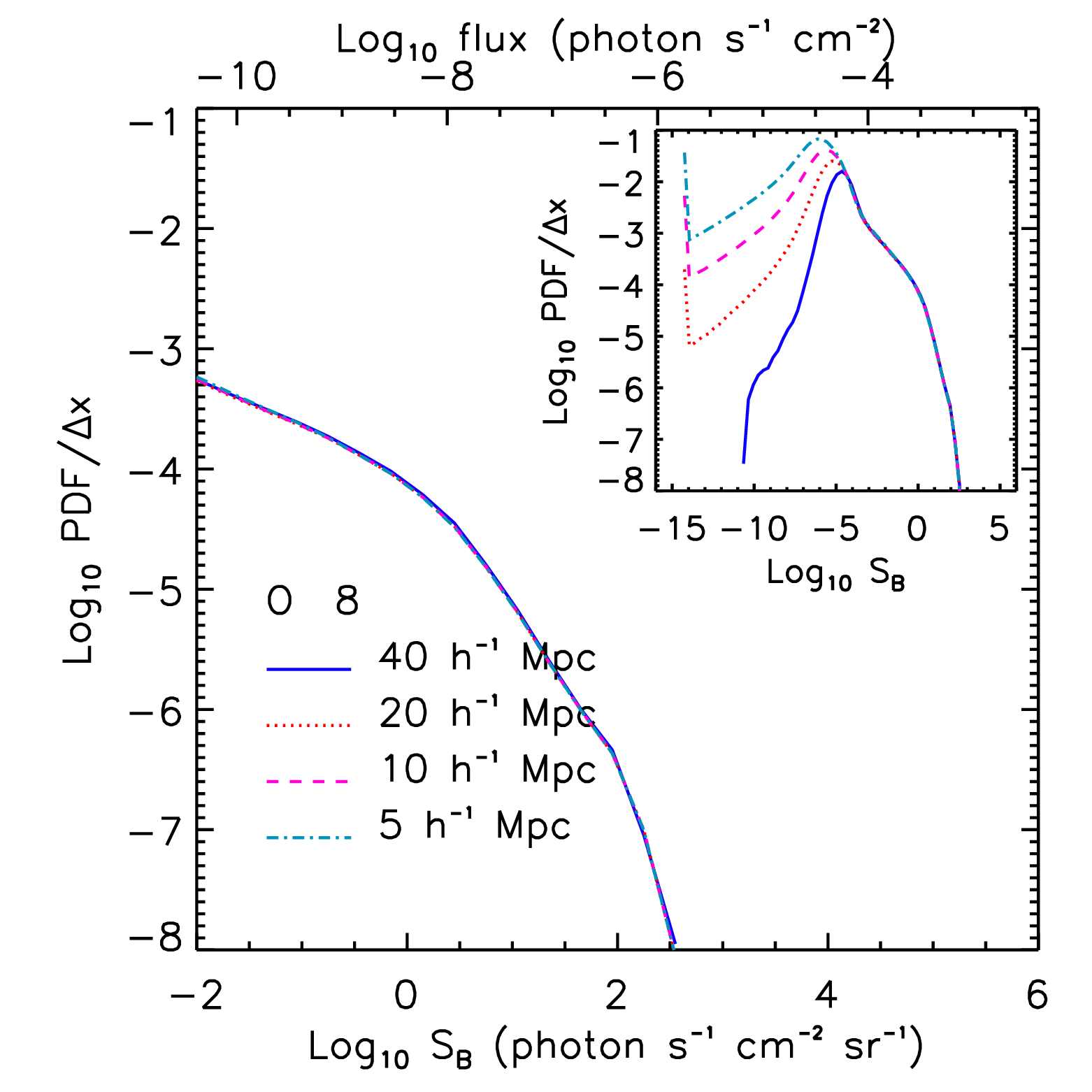}
\caption{The \oviii\ flux PDFs, normalised by the slice thickness, for
  maps with varying slice
  thickness: 40, 20, 10 and 5 \hm\ Mpc. The number of slices used to
  calculate each PDF is 2, 5, 10 and 20, respectively. The test is for
  the \default\ model with box size 100 \hm\ Mpc at $z=0.25$. All maps
  assume the same angular resolution of 15". In the high flux regime
  the PDF is proportional to the slice thickness.}
\label{thick_pdf}
\end{figure}

In this Appendix we describe how varying the thickness of the slices
in which we divide the simulation box affects the flux PDF.
In Fig.~\ref{thick_pdf} we show this by comparing the flux PDFs,
normalised by the slice thickness, for the four
different, evenly-spaced slice thicknesses of 40, 20, 10 and 5
\hm\ Mpc.

In the potentially detectable regime, the results appear to be
independent of the slice thickness when the flux PDF is normalised by
the slice thickness. In other words, the PDF in
the high flux regime is proportional to the slice thickness.  This
is expected if the slice is sufficiently thick to fully contain the
objects giving rise to the emission, but sufficiently thin for the
superposition of multiple structures to be unimportant.

The shape of the PDF at low fluxes, instead, depends on the level of
the flux in low-density regions and on the number of low-density
structures such as filaments that could be found in each slice, which is
dependent on the slice thickness. As a consequence, the actual width
of the PDF at intermediate fluxes scales with the inverse of the slice
thickness and the number of pixels at the lowest fluxes increases for
decreasing slice thickness.

When creating the slices, we neglect the peculiar velocities of
particle, selecting them only according to their position within the
simulation box. This is justified for a large slice thickness, when
the peculiar velocities are small compared with the difference in the
Hubble flow across the slice. 
because the effect of particle displacement is small. However, it may
have a small effect on the predicted fluxes for the 5 \hm\ Mpc slices,
which corresponds to $443~\kms$ at $z=0.25$. 

Throughout this work, we have assumed a slice thickness of 20
\hm\ Mpc, but Fig.~\ref{thick_pdf} demonstrates that the results can
easily be scaled to other slice thicknesses.

\bsp
\label{lastpage}

\end{document}